\documentclass[10pt]{article}
\pdfoutput=1

\newcommand{\QED}{\nobreak\hfill\hbox{$\square$}}

\usepackage{graphicx}
\usepackage{morefloats}

\usepackage{amsmath}
\usepackage{amssymb}

\usepackage{cite}

\usepackage{hyperref}

\usepackage{lineno}

\usepackage{microtype}
\DisableLigatures[f]{encoding = *, family = * }

\usepackage[labelfont=bf,labelsep=period,justification=raggedright]{caption}

\newcommand{\beginsupplement}{%
        \setcounter{table}{0}
        \renewcommand{\thetable}{S\arabic{table}}%
        \setcounter{figure}{2}
        \renewcommand{\thefigure}{S\arabic{figure}}%
     }

\makeatletter
\renewcommand{\@biblabel}[1]{\quad#1.}
\makeatother

\date{}

\begin{document}

\begin{flushleft}
{\Large
\textbf{Weighted Statistical Binning:   enabling
statistically consistent
genome-scale phylogenetic analyses} \\
} 
Md Shamsuzzoha Bayzid$^{1}$,
Siavash Mirarab$^{1}$,
Bastien Boussau$^{2}$,
Tandy Warnow$^{3,\ast}$
\\
\bf{1} Department of Computer Science, University of Texas at Austin, Austin, Texas, USA
\\
\bf{2}Laboratoire de Biom\'etrie et Biologie \'Evolutive, 
Universit\'e de Lyons, France \\
\bf{3} Department of Computer Science, University of Illinois at
Urbana-Champaign, Urbana, IL,   USA\\
$\ast$ E-mail: warnow@illinois.edu
\end{flushleft}

\section*{Abstract}

Because biological processes can result in different loci having different 
evolutionary histories, 
species tree estimation requires multiple loci from across multiple genomes.
While many processes can result in discord between gene trees
and species trees,
incomplete lineage sorting (ILS), modeled by the multi-species coalescent,  
is considered to be a dominant cause for gene tree heterogeneity.
Coalescent-based methods  have been developed to estimate species trees,
many of which operate by combining estimated gene trees, and so are called 
``summary methods".
Because summary methods are generally fast (and much faster
than more complicated coalescent-based methods that
co-estimate gene trees and species trees), they have become
very popular techniques for estimating species trees from multiple loci.
However, recent studies have established that 
summary methods can have reduced accuracy in the presence of
gene tree estimation error, and also that many biological datasets
have substantial gene tree estimation error, so that summary methods may not be 
highly accurate in biologically realistic conditions.
 Mirarab et al. (Science 2014) presented
the ``statistical binning" technique to improve gene tree estimation in multi-locus analyses, and
 showed   that it improved the accuracy of MP-EST,
one of the most popular coalescent-based summary methods. 
Statistical binning, which
uses a simple heuristic to evaluate ``combinability" and then uses the
larger sets of genes to re-calculate gene trees, has good empirical performance, 
but using statistical binning within a phylogenomic pipeline does not have
the desirable property of being \emph{statistically consistent}.
We show that weighting the re-calculated gene trees by the bin sizes makes 
statistical binning
statistically consistent under the multispecies coalescent, and maintains the good 
empirical performance.
Thus, ``weighted statistical binning"
 enables highly 
accurate genome-scale species tree estimation, and is also
statistically consistent under the multi-species coalescent model.
New data used
in this study are available at
DOI: http://dx.doi.org/10.6084/m9.figshare.1411146, and
the software is 
available at https://github.com/smirarab/binning.



\section*{Introduction}
The estimation of phylogenetic
trees, whether of  individual loci (so called ``gene trees") or
at the genome-level  (species trees), is a
basic step in many biological analyses
\cite{Eisen1998}.
However, estimating gene trees and species trees with
high accuracy is difficult for many reasons, including
computational issues (nearly all problems are NP-hard)
and dataset issues.
For example,  while highly accurate
gene trees can be computed for some loci, 
when a locus has 
limited \emph{phylogenetic
signal} (e.g., its sequences are too short, or it
evolves too slowly),
its gene tree may only be estimated
with partial accuracy. 
Species tree estimation is also difficult,
because different loci can have different phylogenetic trees, 
a phenomenon that occurs due to several different
biological processes.
In particular, many groups of species evolve with rapid
speciation events, a process that is likely to produce
conflict between gene trees and species trees due to
\emph{incomplete lineage sorting} (ILS) \cite{maddison,degnan-rosenberg,edwards,Rosenberg2013}.
Furthermore, when ILS occurs, standard methods for estimating species trees, such as concatenation (which combines sequence
alignments from different loci into a single ``supermatrix", and then computes a tree on the supermatrix)
and consensus methods, can be statistically inconsistent \cite{RochSteel-inconsistent,consensus-inconsistent}, and produce highly supported but incorrect trees \cite{kubatko-degnan-2007}. Because these standard methods for estimating species trees from multiple loci can be positively
misleading in the presence of gene tree heterogeneity due to ILS, statistical methods (e.g., \cite{stells,stem,heled-drummond,DeGiorgioDegnan,star})
have been developed to estimate the species tree assuming all gene tree 
heterogeneity is due to ILS and, in particular, not to poor phylogenetic signal.


Here we describe one of the recent  approaches for estimating the species tree from
a set of 
multiple sequence alignments, one for each of $p$ different loci on a set $S$ of $n$
species.
We will assume that the input sequence data are generated under a multi-step process, which we
now define:

\vspace{.1in}
\noindent
Definition 1:
Under the {\bf GTR+MSC} model,  gene trees evolve
within a species tree under the multi-species coalescent (MSC) model,
and then sequences evolve down each gene tree under the General
Time Reversible (GTR) model \cite{gtr}. The different gene trees
are equipped with their own GTR model parameters, and so
the tree topologies, $4 \times 4$ substitution matrices, and
gene tree
branch lengths
can differ between the different genes. 
\vspace{.1in}

Thus, under the GTR+MSC model, a method for estimating the
species tree will begin with the sets of sequences for 
the different loci, and then infer the species tree.
There are many different types of methods to
estimate species trees from sets of sequence
alignments for multiple loci, 
and we will refer to all of these methods as ``phylogenomic
pipelines".

\vspace{.1in}
\noindent
Definition 2: 
We will say that a phylogenomic pipeline 
is {\bf statistically consistent} under
the GTR+MSC model if,
as the number $p$ of loci and the number $k$ of sites in the
sequence alignment for each locus both increase to infinity, then the estimated
species tree converges in probability to the true species tree.
\vspace{.1in}

There are many  phylogenomic
pipelines  that are statistically consistent under
the GTR+MSC model, but in this study we focus on pipelines that
operate by first estimating gene trees and then
combining these estimated gene trees using a summary method. 
More specifically, we will restrict the discussion to
pipelines that use ``coalescent-based" summary methods, 
as follows:

\vspace{.1in}
\noindent
Definition 3: 
A {\bf coalescent-based summary method} is a 
method that estimates the species tree by combining
gene trees, and which
converges in probability to the true species tree as the
number of true gene trees sampled from the distribution
defined by the species tree increases.
\vspace{.1in}

Examples of coalescent-based summary methods include
MP-EST \cite{LYE-mpest}, ASTRAL \cite{astral,astral2}, STAR \cite{star}
and NJst \cite{njst}.
Coalescent-based 
analyses of biological datasets typically
use this kind of pipeline, since they can be computationally
more efficient than other types of coalescent-based analyses 
(for example, methods like *BEAST \cite{Heled2010}
that co-estimate the gene trees and species tree).

Thus, we focus the discussion in this study on
phylogenomic pipelines that have the following 
basic structure:
\begin{itemize}
\item Step 1:  a gene tree is estimated for each
locus
\item Step 2:  the gene trees are combined
into a species tree using a  coalescent-based
summary method. 
\end{itemize}


While many studies have 
explored the statistical properties of 
coalescent-based summary methods given true gene trees, here
we focus on their use when the input
is a set of sequence alignments for multiple loci.
In this context, 
proofs of statistical consistency for GTR+MSC phylogenomic
pipelines 
have taken the following form  \cite{RochWarnow}.
First, gene trees are estimated using a statistically 
consistent method, such as GTR maximum likelihood, 
and we assume that the
sequences for each locus are long enough that 
the true gene tree is computed with high probability.
Then, 
the species tree is estimated 
using a coalescent-based summary method. Thus,
the proofs of statistical consistency under
the GTR+MSC model for
common coalescent-based summary methods (e.g., 
ASTRAL, MP-EST, NJst, etc.) have explicitly or
implicitly assumed
that true gene trees are given as input to the summary
method.


Little mathematical theory has been proven about the
impact of gene tree estimation error on
coalescent-based summary methods; for example,
it is not known
whether standard summary methods will converge
to the true species tree
given a large enough number of gene trees, if
each gene tree is estimated from bounded
length sequences \cite{RochWarnow}.
Furthermore,  empirical 
studies suggest that summary  methods are
impacted by gene tree estimation error, and
can produce less accurate estimated species trees than
concatenation when gene tree estimation error is high enough
(see \cite{naive-binning,Patel2013,MirarabStatBinning,GatesyMPE2014,RochWarnow} for
examples of these studies on summary methods and further discussion).
In a genome-scale analysis,  it is unlikely that all the loci
will have substantial phylogenetic signal, and so this
vulnerability to gene tree estimation 
error means that coalescent-based
summary 
methods may not be highly accurate techniques for
estimating species trees from genome-scale data.
This is particularly problematic when the sequences for each locus are kept short to diminish the probability of intra-locus recombination (which violates the assumptions of the multi-species model), since short sequences will tend to have insufficient phylogenetic signal to provide full resolution of the gene trees; see \cite{Patel2013,GatesyMPE2014,LanierKnowles2012} for discussion about this important issue.

In \cite{MirarabStatBinning}, we developed a technique
we called ``Statistical Binning" to improve species tree
estimation using phylogenomic pipelines based
on coalescent-based summary methods.  
Statistical binning partitions the genes into sets
based on a heuristic to evaluate ``combinability", 
concatenates the gene sequence alignments within each set into a
``supergene alignment", and then estimates a 
``supergene tree" on
the supergene alignment using a fully partitioned maximum likelihood analysis.
The newly estimated supergene trees are then
used by the preferred coalescent-based summary method to compute a species tree
on the dataset.
As shown in \cite{MirarabStatBinning}, statistical binning 
improved
the estimation of gene trees and gene tree distributions, and this 
resulted in improved estimates of the species tree topology and
branch lengths when species trees were computed using
MP-EST with multi-locus bootstrapping (MLBS). 
Furthermore, when used with statistical binning, MP-EST
was almost always at least as accurate 
as concatenation (more accurate than concatenation when
the ILS level is high, and only less accurate than concatenation
for very low levels of ILS).
Finally, 
MP-EST used with
statistical binning was used to compute
a species tree on the
avian phylogenomic dataset, and this
``MP-EST*" tree was nearly identical to the
concatenation analysis we obtained; the MP-EST* 
and concatenation trees
were presented in \cite{JarvisScience2014} as the two major hypotheses
for the avian 
phylogeny.

Thus, Mirarab {\em et al.} proposed a new type of
phylogenomic species tree estimation pipeline that 
has four steps instead of two (where the
extra two steps 
are partitioning the genes into bins based on perceived ``incompatibility", and 
computing supergene trees for each bin using 
a fully partitioned maximum likelihood analysis). 
This pipeline, which we referred to as ``statistical binning", 
showed very promising results when used with MP-EST. However,  we did
not address the theoretical properties of these pipelines, 
we only examined model trees with 37 or more species, and
we only analyzed one coalescent-based summary method, MP-EST.

In this paper, we report on an extended evaluation of
statistical binning.
Specifically,
\begin{itemize}
\item 
We provide a proof of statistical inconsistency under
GTR+MSC for pipelines based on 
the original protocol for statistical binning presented in \cite{MirarabStatBinning}.
\item 
We describe a variant of
statistical binning that we call ``weighted statistical binning", and
provide a proof of statistical consistency under GTR+MSC for 
pipelines based on weighted statistical binning.
\item
We evaluate the impact of 
statistical binning (both weighted statistical
binning and the original  unweighted statistical binning technique) on biological 
and simulated datasets under the GTR+MSC model.
 We   examine  pipelines using two 
coalescent-based summary methods,  ASTRAL and MP-EST. 
We include
results on simulated and biological datasets studied in \cite{MirarabStatBinning}, 
and also on additional simulated
datasets with 10 and 15 taxa. 
\end{itemize} 

The study shows that weighted and unweighted statistical binning
have very similar results across most datasets,  and also that both 
ASTRAL and MP-EST tend to improve in accuracy when used with binning.
However, there was one condition (characterized by
a very high level of ILS, low average bootstrap support for  the
gene trees,  and only ten species) 
in which statistical binning reduced accuracy for both
MP-EST and ASTRAL. 
Thus, this study shows that binning is often beneficial, but
also that there are some conditions under which binning can 
increase rather than decrease species tree error. 
Finally, we  conclude with suggestions for further research.

\subsection*{Weighted Statistical Binning}

The statistical binning technique presented in 
\cite{MirarabStatBinning} operates 
as follows.
The input is
a multiple sequence alignment 
  on each of $p$ given genes, 
and a user-specified ``threshold support" value $B <1$.
The role of the threshold $B$ is to specify which branches in the
gene trees are considered reliable, and which ones have support
that is so low that the branches may be due to 
estimation error.
Therefore, if the trees on two genes
differ only in their low support edges, the differences are
considered potentially consistent with estimation error,
and the two genes are considered ``combinable" or ``compatible".

Statistical binning computes maximum
likelihood (ML) gene trees and bootstrap
support on the branches for 
each gene, and then
uses a simple heuristic
based on bootstrap support values
so that
two genes can only be in the same bin if their ML gene
trees do not have conflicting branches, each with bootstrap support of
at least $B$.
This is the combinability test, so that two genes are not 
considered
combinable
if they have highly supported conflicting branches, and otherwise are
considered combinable.
(Equivalently, two genes are combinable if their ML gene trees, after
collapsing all branches with support less than $B$, share a common refinement.)
Finally, because pairwise compatibility ensures setwise compatibility
\cite{gusfield},
if a set of gene trees \emph{can} be all put in the same bin,
then there is a tree that combines all the highly supported
branches in any of the trees in the set.

\paragraph{Computing and using the incompatibility graph to bin the genes. }
The first step in statistical binning
creates a graph based on the input, 
and uses a graph-theoretic optimization to bin the genes into subsets. 
Each gene is represented by a single vertex in the graph, and
an edge is placed between two genes if their gene trees are not combinable,
based on the heuristic described above.
Determining if two genes are combinable can be computed 
in linear time \cite{Warnow-treecompat}, and so this graph,
which we call the incompatibility graph, can be computed in time linear in 
the number of taxa and quadratic in the number of genes.

Since longer sequences tend to produce more
accurate gene trees,   
having the  bins be as
large as possible is desirable; this is 
accomplished indirectly by seeking a coloring with as few colors as
possible (i.e., a minimum vertex coloring), which is an NP-hard
problem \cite{Karp1972}.
However, summary methods, such as MP-EST, use the distribution of the gene trees to estimate the species tree. Assuming gene tree reconstruction error only results in low-support branches, binning the genes so that the bins have nearly the same size 
means  that the supergene tree frequency will be close 
to the true gene tree distribution (assuming that
binning combines genes with the same tree, and that
we can compute correct supergene trees). 
Note also that
with such a constraint, frequent true gene tree topologies 
will be represented in several bins, while each of the rarest 
gene trees will be represented in a smaller
number of bins (and perhaps in only one bin).
Therefore, the objective is a coloring of the vertices, using a small
number of colors, so that every color class contains about the same number
of colors. To achieve such a coloring,
\cite{MirarabStatBinning} modified 
the Br\'elaz heuristic \cite{brelaz1979new} for
minimum vertex coloring,
so that during the greedy coloring, a node is added to the smallest bin for which
it has no conflicts.
This coloring produces a partitioning of 
the vertices of the graph into subsets based on the color classes;
thus, all vertices with the same color are in the same bin.

\paragraph{Computing a supergene tree for each bin. }
Once the vertex coloring is computed,  the genes in a given
color class form a bin, and their alignments are concatenated into
a supergene alignment. Then, a maximum likelihood tree is computed
(perhaps with bootstrapping)
on each supergene alignment. 
For estimating supergenes, we use a {\em fully partitioned
analysis} where each gene is assigned a separate partition, 
and all numeric model parameters are allowed to differ between partitions. 
We call the
trees that are computed on
the supergene alignments   ``supergene trees".
Because using a fully partitioned analysis is key to the theoretical
guarantees of statistical binning, we specifically discuss this step
in the pipeline.

Concatenated ML analyses of alignments from different loci can be
performed in many ways, but their theoretical properties
depend on the details of how they are performed, and in
particular whether they are performed using
an unpartitioned analysis, or a partitioned analysis. 
In an unpartitioned analysis, 
all the sites in the concatenated alignment 
are assumed to evolve down a single model tree
(i.e., topology and numeric parameters), and
the model tree maximizing the likelihood is sought for the matrix. 
In contrast, fully partitioned analyses of concatenated
alignments assume that the different loci all evolve
down the same tree topology, but  allow the different parts within
the concatenated alignment to have different values for all
of the numeric parameters of the model. In
the context of the GTR model, 
 a fully partitioned maximum likelihood analysis would
allow each locus to have its own $4 \times 4$ substitution
matrix and gene tree branch lengths. 
Thus, if there are $10$ loci within the concatenated alignment,
a single tree topology is returned, but also ten different 
lengths for each branch, and ten different $4 \times 4$ substitution
matrices. 
Fully partitioned and unpartitioned maximum likelihood analyses can 
result in different trees, and these analyses have very
different theoretical properties; see
the example provided
in the Methods section, below.

\begin{figure}
\centering
\begin{tabular}{c}
\includegraphics[width=0.90\textwidth]{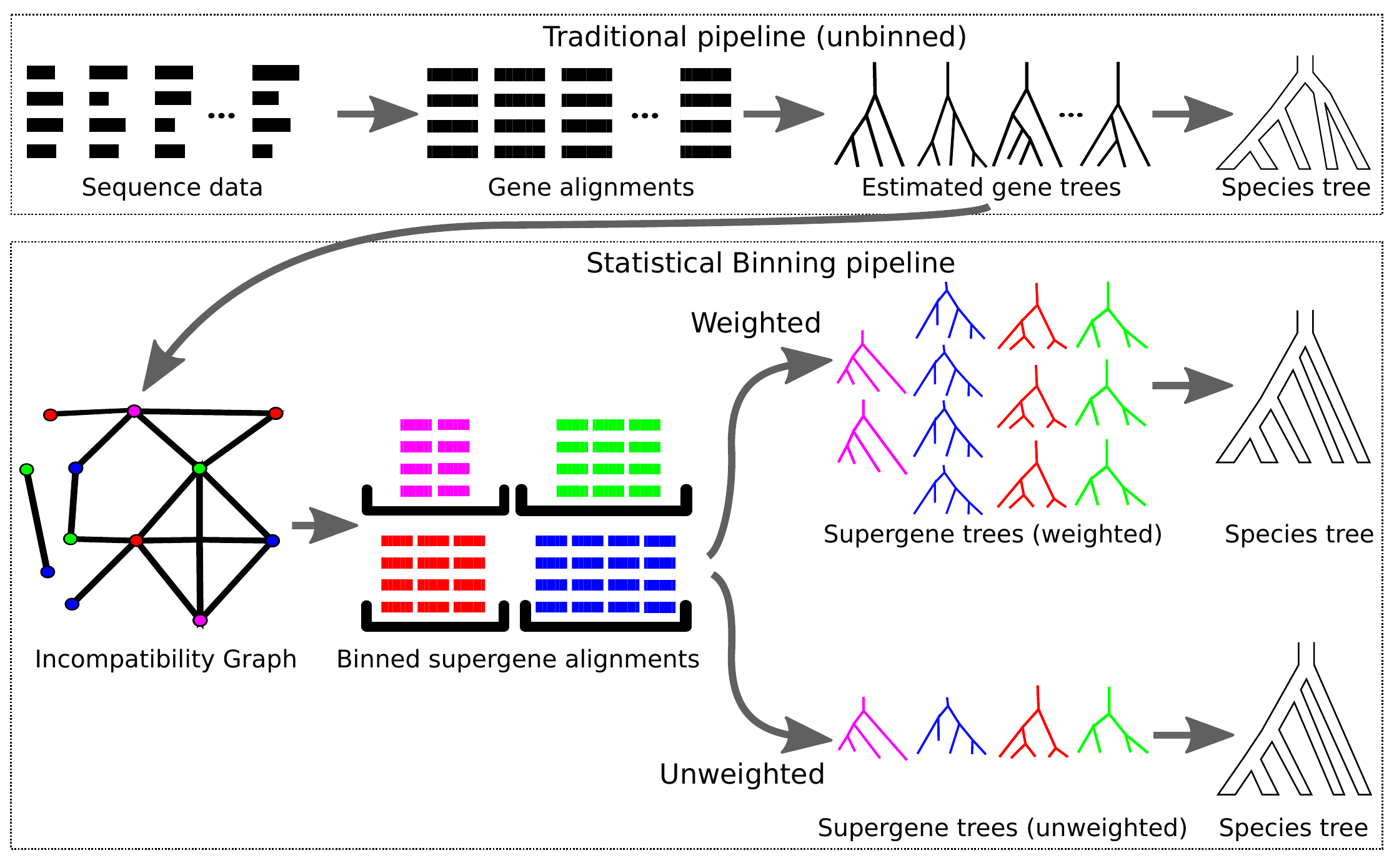} 
\end{tabular}
\caption{
{\bf Pipeline for unbinned analyses, unweighted statistical binning,
and weighted statistical binning.}
The input to the pipeline is a set of sequences for different loci across different species. 
In the traditional pipeline, a multiple sequence alignment 
and gene tree is computed for each locus, and then these are given
to the preferred coalescent-based summary method, and a species tree is returned.
In the statistical binning pipeline, the estimated gene trees are used to compute
an incompatibility graph, where each vertex represents a gene, and an edge between two genes indicates that the
differences between the trees for these genes is considered significant (based on the bootstrap
support of the conflicting edges between the trees).
The vertices of the graph are then assigned colors, based on a heuristic for balanced minimum vertex coloring,
so that no edge connects two vertices of the same color.  The vertices with a given color are put into a bin, and
the sequence alignments for the genes in a  bin are combined into a supergene alignment. 
A (supergene) tree is
then computed for each supergene alignment using a fully partitioned
analysis.  In the unweighted binning approach (presented in
\cite{MirarabStatBinning}), these supergene trees are then given to the preferred summary method, and a species
tree is returned.  In the weighted binning approach presented here, each
supergene tree is repeated as many times as the number of
genes in its bin, and this larger set is then
given to the preferred summary method.
}
\label{fig1}
\end{figure}

\paragraph{Applying summary methods to the supergene trees. }
The supergene trees are used by a coalescent-based summary
method (e.g., MP-EST) to estimate the species tree.
In other words, by recalculating the
gene trees, statistical binning changes the input to the coalescent-based summary method.
Hence, statistical binning is a technique to
re-estimate gene trees used within the coalescent-based
pipeline for 
species tree estimation, as shown in Fig.~\ref{fig1}.

\paragraph{Theoretical properties of pipelines based on statistical binning. }
Theorem 3 shows 
that the use of statistical binning within a phylogenomic pipeline
is not statistically 
consistent under the GTR+MSC model.
The failure to be statistically consistent occurs 
because we do not {\em require} that the bins be equally sized;
hence, the distribution on supergene trees 
can be different from
the true gene tree distribution.

However, a simple variation of the technique, which was suggested
in \cite{MirarabStatBinning},  corrects this problem.
We keep the first step of statistical binning the same (i.e., we 
compute the same incompatibility graph and then use 
the same heuristic for balanced minimum vertex coloring), and
we compute the same set of supergene trees. However, 
at this point
we replicate every supergene tree so that it appears as many times
as the number of genes in its bin.  For example, if we begin with $100$ genes,
and obtain $20$ bins, then the original statistical binning
technique would produce $20$ supergene trees that would be given to MP-EST to
analyze. In this modified technique, if we begin with $100$ genes, we end up
with $100$ supergene trees (although some supergene trees will be identical).
We call this technique ``weighted statistical binning", and refer
to the original technique proposed in \cite{MirarabStatBinning} as
``unweighted statistical binning".
We prove that the use of weighted statistical binning
is statistically consistent
in Theorem 2. 

Fig.~\ref{fig1}  describes the three possible pipelines (unbinned, unweighted binned,
and weighted binned) for use with a summary method.
In the unbinned analysis, each gene is analyzed independently,
a gene tree is estimated for each gene,
and then a summary method, such as MP-EST, uses the gene trees to 
estimate the species tree.
In both the weighted and unweighted binned analyses, the gene trees 
are computed independently, and then the
incompatibility  graph is formed with one vertex for
each gene. In the shown example, there are 12 genes, and so the graph has
12 vertices. The 12 vertices of the incompatibility graph
are then assigned colors, with two vertices colored purple,
three vertices colored green, three vertices colored red, and four 
vertices colored blue. Note that no two vertices of the same color
have an edge between them.
For each color class, the sequence alignments for the associated
genes are concatenated into one long supergene alignment, and a supergene
tree is computed on the
supergene alignment using a fully partitioned 
maximum likelihood analysis. After this point, the weighted and unweighted binning
methods have different strategies. In the unweighted binning method,
exactly one copy of each supergene tree is given as input to 
the summary method, but in the weighted binning method multiple copies of
the supergene trees are given as input.
Hence, in this example, MP-EST analyzes only four supergene trees in the
unweighted binning pipeline, but it analyzes 12 supergene trees
in the weighted binning pipeline.

By design,
if the bin sizes are exactly the same, then
the statistical binning pipelines produced using
weighted and unweighted statistical binning produce the same
results; hence, these two approaches can only produce
different results when the binning is unbalanced.

\subsection*{Experimental Study}

\paragraph{Datasets. }
We use 
the avian and mammalian simulated datasets studied in
\cite{MirarabStatBinning} (each based on 
MP-EST analyses of 
biological datasets, and having at least 37 taxa) and two other collections of simulated
datasets with 10 and 15 taxa.
The simulated datasets range from moderately low ILS (the lowest
ILS mammalian condition) to extremely
high ILS conditions (the higher ILS 10-taxon and 15-taxon
model conditions), and range in terms of average gene tree
bootstrap support (from very low to moderately high). 
Thus, the simulated datasets provide a range of conditions
in which we explore the impact of statistical binning.
We also analyzed two biological
datasets (a 48-species avian dataset and a 37-species mammalian
dataset) studied in 
\cite{MirarabStatBinning}. 

\vspace{.1in}
\noindent
We used biologically-based simulated datasets that were 
studied in \cite{MirarabStatBinning}, and are based on
species trees estimated using MP-EST on the avian dataset
of \cite{JarvisScience2014} and 
the mammalian
dataset
of \cite{Song2012}.
In the avian simulation, the markers vary in sequence length
(250bp, 500bp, 1000bp, and 1500bp)
in order to produce bootstrap support values similar to those
we observed
in the biological dataset.
In the mammalian simulation, we again explored the impact of
phylogenetic signal by varying the sequence length (250bp, 500bp, and 1000bp)
for the markers.
In both cases, three levels of ILS are simulated by multiplying or 
dividing all internal branch lengths
in the model species tree by two, 
and we also explore various numbers of genes. 
The mammalian datasets range in ILS level from relatively low
(18\% average distance between true gene trees and the species tree)
for the 2X branch length level to relatively high (54\% average distance between
true gene trees and the species tree) for the 0.5X branch length
level, and the average bootstrap support on the estimated gene trees
ranges from low (43\%) for the shorter (250bp) sequences to
moderately high (79\%) for the longest (1000bp) sequences. 
The avian datasets have higher ILS levels than the
mammalian datasets, and range from 
moderate 
(35\% average distance between true gene trees and the species
tree) for the 2X branch length condition to 
high (59\% average distance between true gene trees and the 
species tree) for the 0.5X branch length condition.
The estimated gene trees
range in average bootstrap support from very low (27\%) for the shortest
(250bp) sequences to moderate (60\%) for the longest (1500bp)
sequences.

\vspace{.1in}
\noindent
We also used a 15-taxon model species tree with a caterpillar-like
(also known as a 
pectinate, or ladder-like) topology, which has 12 short internal branches
(0.1 in coalescence units)
in succession, a condition that gives rise to high levels of ILS \cite{kubatko-degnan-2007,Rosenberg2013-agt}.
Ultrametric gene trees were simulated down this tree using 
the multi-species coalescent process (see Methods). 
Unlike the biologically-based model conditions,
no transformations of branch lengths were used,
and therefore, gene trees follow a strict molecular clock. 
Sequence data were simulated down each gene tree, and
we built four model conditions by trimming gene data to 100 or 1000 sites, 
and by using 100 or 1000 genes. 
This dataset is very homogeneous since all 10 replicates we simulated are
based on the same species tree, and gene trees differ
in topology and branch length only due to the coalescence process.
The 15-taxon datasets have very high ILS levels (82\%
average topological distance between true gene trees
and the species tree), and so represent a rather
extreme condition. The gene trees
estimated on the shorter sequences 
(100bp)   had only 35\% average bootstrap support, 
and the combination of very high ILS and very low
average bootstrap support represents a very challenging condition.
Gene trees estimated on the longer sequences have better average
bootstrap support (70\%), and so represent a somewhat easier condition.

\vspace{.1in}
\noindent
 We also 
generated 10-taxon simulated
datasets using simPhy \cite{simphy}. In
this simulation protocol, we simulated a different species tree
for each replicate, and simulated 200 gene trees for each species tree using the multi-species coalescent process. We simulated two model conditions, one
 with very high ILS and another with  somewhat lower (but still high) ILS. 
 The simPhy procedure 
 uses a host of various distributions to make the gene trees heterogeneous in 
various aspects, such as sequence
lengths, deviation of branch lengths from the strict molecular clock, 
and rate variation across different genes.
We used these gene trees to simulate sequence data with 100 sites using Indelible \cite{indelible}. 
Therefore, our 10-taxon datasets are very heterogeneous: different replicates have
different species trees, and within each replicate, various genes have 
different rates of evolution. 
The ILS levels of the 10-taxon datasets range from
moderately high (40\% average distance
from true gene trees to the species tree) for the ``lower ILS" condition
to extremely high (84\% average distance)
for the ``higher ILS" condition. 
The average bootstrap support on the estimated gene trees ranged from 37\% for
the higher ILS condition to 45\% for the lower ILS condition, and so
both have very poor average bootstrap support.
Thus, the 10-taxon and the 15-taxon datasets with
short sequences represent the hardest model conditions in that they have
very high ILS and very low average bootstrap support.

\vspace{.1in} \noindent 
The simulated datasets we studied varied in many respects
(sequence length per locus, whether the sequence
evolution is ultrametric or not, and the ILS level). 
Table \ref{table:model-condition}
presents data about the ILS level, as reflected in the
average topological distance between the true gene trees and the
true species tree.
Note that two of the model conditions (the 10-taxon higher
ILS and 15-taxon datasets) have extremely
high ILS, reflected in average topological distances
between the true gene trees and the species tree. 
In fact, most of the model conditions have high
ILS levels (with 30\% or more average topological
distance between the true gene trees and the species tree),
and only one model condition has low levels of
ILS (the mammalian datasets with 2X branch lengths, which
have 18\% average topological distance between the
true gene trees and true species tree). 
It is likely that the ``1X" ILS levels for the mammalian
and avian simulated datasets are larger than the
ILS levels for the respective biological datasets, since the model
trees
that were used to generate these data
are based on MP-EST analyses
of the datasets, and results in \cite{MirarabStatBinning}
suggest that MP-EST estimations tend to 
under-estimate branch lengths, and hence inflate
estimated
ILS levels.

\begin{table}[htbp]
\centering
\begin{tabular}{llcc}
\hline
Dataset & ILS level & Discordance (\%)\\ \hline
Avian     & 2X & 35 \\
Avian & 1X & 47  \\
Avian      & 0.5X & 59  \\ 
Mammalian          & 2X & 18 \\
Mammalian & 1X & 32 \\
Mammalian          & 0.5X & 54  \\ 
10-taxon & Lower ILS & 40  \\
10-taxon         & Higher ILS & 84  \\ 
15-taxon & High ILS & 82 & \\ 
\end{tabular}
\caption{\textbf{Topological discordance between true gene trees and
true species tree}. For each
collection of simulated datasets (defined by the
type of simulation and the ILS level), we
show  the average topological distance between true gene
trees and the species tree. 
}
\label{table:model-condition}
\end{table}

\paragraph{Methods. }
We computed coalescent-based species
trees using summary methods with MLBS gene trees
in three ways: without binning, 
with weighted statistical binning and with
unweighted statistical binning.
Our main focus is on MP-EST, but we explore
results with ASTRAL 
 on a subset of the data.
ASTRAL estimates species trees given
unrooted gene trees, and can analyze very large datasets (such as the
plant transcriptome dataset with approximately 100 species and
600 loci \cite{1kp});
hence, ASTRAL can analyze larger datasets than
MP-EST, and so understanding the impact
of binning on ASTRAL's accuracy is of
practical importance.

We perform statistical binning using both weighted and unweighted 
pipelines and using two support thresholds ($B$): 50\% and 75\%. 
Due to the extremely large computational effort involved, 
on our two large biologically-based simulated datasets, 
we explore one 
threshold for most of our results;
we follow the protocol used in \cite{MirarabStatBinning}
and set $B=50\%$ for the avian datasets, and $B=75\%$ for the mammalian datasets.
However, we also study the impact of $B$ on one model condition for 
avian and mammalian datasets. 

We compute gene trees and concatenation 
species trees using RAxML \cite{raxml} maximum likelihood.
For estimating supergene trees, we use fully partitioned RAxML analyses (using the $-M$ option
to vary branch lengths across genes)
for smaller (10- and 15-taxon) simulated datasets and for all biological 
analyses. However, since partitioned 
analyses are expensive, we use unpartitioned analyses 
to compute supergene trees 
for our studies on the avian and mammalian simulated datasets
(because these studies are very extensive).
We compare results using coalescent-based summary methods
to concatenation, also  using unpartitioned maximum likelihood. 
Note that the binned methods and the concatenation analysis would
potentially become more accurate if fully partitioned analyses were
employed.

\paragraph{Measurements. }

For the simulated datasets, we explore species tree accuracy
with respect to the true (model) species tree topology
(the missing branch rate, or false negative rate (FN))
and branch lengths, and also examine the branch support
of both true positive and false positive branches.
We also explore the error in the estimated
gene trees and  gene tree distribution estimated
using binning (weighted and unweighted), compared to unbinned analyses.
We analyze these simulated datasets using
weighted statistical binning with MP-EST and
ASTRAL, to determine if there are 
differences between
weighted and unweighted statistical binning. 
Since ASTRAL does not produce branch lengths, we only use MP-EST to evaluate
branch length estimation. 
In addition, we
examine the bootstrap support on the branches of
estimated species trees produced using MP-EST, as false positive edges
that have low support are not as deleterious as
false positive edges with high support.
The bootstrap support of estimated species trees
was not studied in \cite{MirarabStatBinning},
and so this study provides the first analysis of
bootstrap support for MP-EST on these datasets, 
as well as of  the impact 
of binning on bootstrap support values.

These aspects of phylogenomic estimation are important
for different reasons. 
Species tree topologies 
indicate which species are more closely related to each other than
to
others, and so estimating 
accurate species tree topologies is the most important
aspect of
phylogenomic estimation. However, the improvement in species tree (coalescent-unit) branch length estimation is
also biologically relevant, since these lengths are related to effective population sizes and generation times of ancestral species, 
and are also used to estimate the amount of ILS in the data. 
Bootstrap support is important, since low support
branches are often ignored,  but high support
branches are generally assumed to be correct; hence,
understanding  whether a method returns high support
for false positive branches (indicating incorrect relations
within a tree) is particularly important.
Improvements in estimating the gene tree distribution 
matter because the accuracy of summary methods depends
on an input that captures the correct gene tree distribution.


For the biological datasets,
we compare estimated species trees to the literature
for each dataset, focusing on whether the estimated species
tree violates known subgroups for the phylogeny.

\section*{Results and Discussion}

\subsection*{Biologically-based simulated datasets}


\subsubsection*{Gene tree error and gene tree distribution error on avian simulated datasets}
We evaluated the impact of statistical binning on
gene tree estimation error for the 1X (default ILS) 
model condition, with sequence lengths varying
from 250bp to 1500bp.
At the shorter sequence lengths, gene
tree estimation error was reduced substantially
(from 79\% to 57\% for 250bp, and from 69\% to 57\% for 500bp)
(S1 Table). Gene tree estimation error was reduced
slightly at 1000bp (from 55\% to 51\%) and even less at
1500bp (from 46\% to 45\%). 
Hence, when gene tree estimation error is high due to
insufficient sequence length, then binning
reduces gene tree estimation error, but binning has
little impact on gene tree estimation error when the sequences
are long enough.

\begin{figure}
\centering
\begin{tabular}{c}
\includegraphics[width=0.95\textwidth]{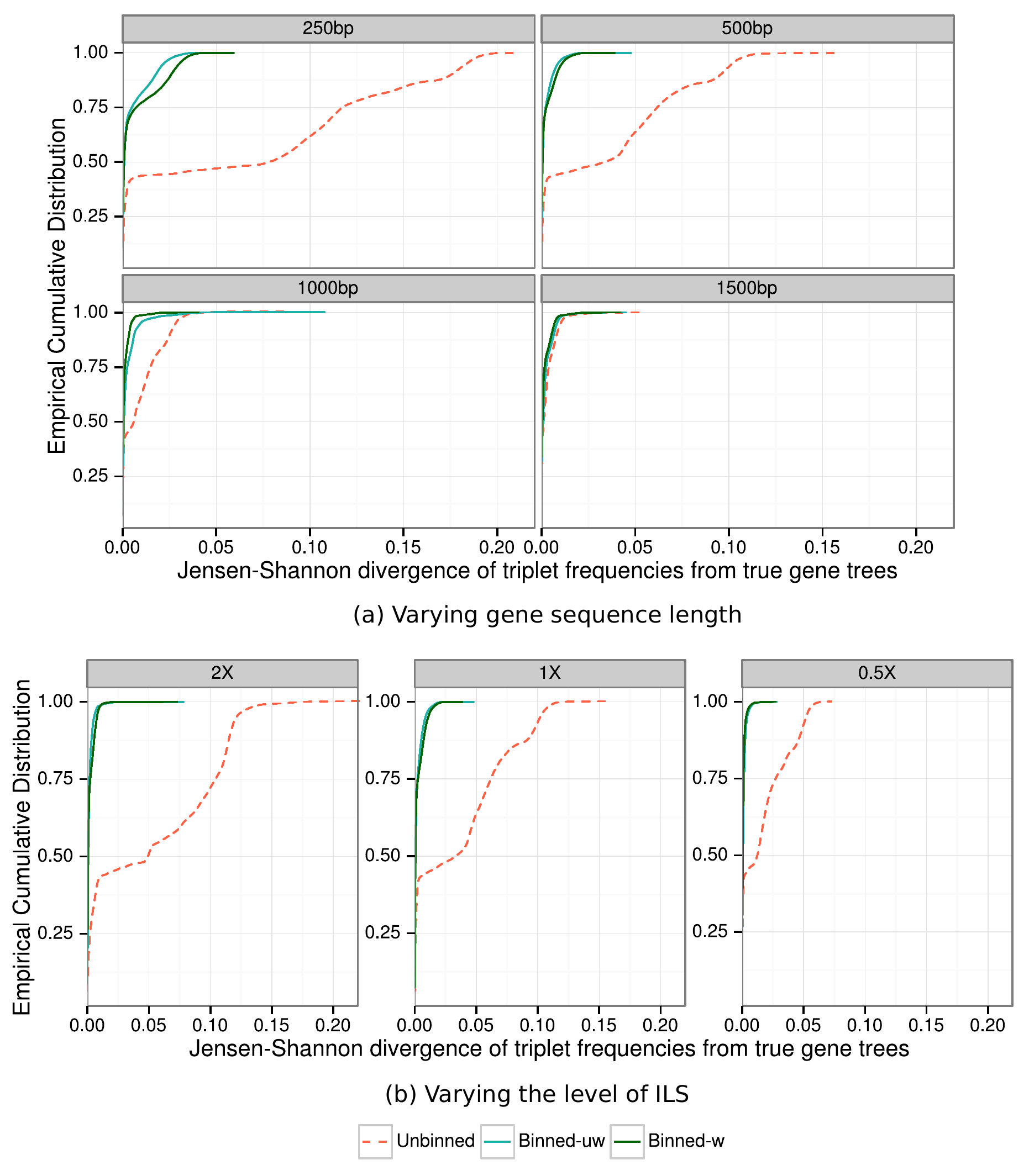} 
\end{tabular}
\caption{
{\bf Divergence of estimated gene tree 
(triplet) distributions from true gene tree
distributions for MP-EST analyses of simulated avian datasets.}
In (a), we vary the gene sequence length (250bp genes have
the highest error, and 1500bp has the lowest error)
and explore 1000 genes under default ILS levels, 
and in (b) we vary
the amount of ILS and fix the number of genes to 1000 and sequence length 
to 500bp. 
True triplet frequencies are estimated based on true gene trees for each 
of the $n \choose 3$ possible triplets, where $n$ is the number of species. 
Similarly, triplet frequencies are calculated from estimated gene/supergene trees. 
For each of these $n \choose 3$ triplets, we calculate the 
Jensen-Shannon divergence of the estimated triplet distribution from the true gene tree triplet 
distribution. We show the empirical cumulative distribution of these divergence scores. 
The empirical cumulative distribution shows 
the percentage of the triplets that are diverged from the true triplet distribution at or below the specified divergence level. 
Results are shown for 10 replicates. We used 50\% bootstrap support threshold for binning, and estimated the supergene trees using RAxML with unpartitioned analyses.  
}
\label{fig-triplet-avian}
\end{figure}

We measure the error in estimated gene tree distributions 
using the deviation of triplet frequencies from the 
triplet frequency distribution computed 
using  the  true gene trees (see Materials and Methods).
We express these results
using a cumulative distribution over all possible triplets and all replicates;
hence, if a curve for one method lies above
the curve for another method, then the first method
strictly improves on the second method with respect
to estimating the gene tree distribution.
In Fig.~\ref{fig-triplet-avian}(a) we show results
for 1000 avian genes under default ILS levels, 
as we vary the sequence length.
In Fig.~\ref{fig-triplet-avian}(b) we show
results with 
1000 genes of length 500bp, varying the ILS level.
In both cases, both weighted and unweighted binning
are nearly identical.
Weighted and unweighted binning also show nearly identical
gene tree distribution errors under other conditions (see 
S1 Fig.). 
Binning improves the accuracy of estimated gene tree
distributions in general, but
not for the longest sequences (1500bp).
Also, the improvement over unbinned analyses was highest for the
lowest ILS level (2X species tree branch lengths), but was high
even for the highest ILS level we explored.

\subsubsection*{Species tree estimation error on avian simulated datasets}
Fig.~\ref{fig-fn-avian} shows results for species tree topology estimation error
for analyses of avian genes of different length under
the default ILS level using MP-EST and ASTRAL,
for varying number of 500bp genes with default ILS using MP-EST,
and  for 1000 genes of 500bp with varying ILS using MP-EST. 
Weighted and unweighted statistical binning are essentially identical
for both MP-EST and ASTRAL
(no statistically significant differences were observed
according to a two-way ANOVA test; see Tables \ref{table-test-mpest} and \ref{table-test-astral}),
and both reduce species tree
estimation error
compared to unbinned analyses (differences were always 
statistically significant with p $<$ 0.001; see Tables \ref{table-test-mpest} and \ref{table-test-astral}).

\begin{table}[h]
\begin{small}
\begin{tabular}{llccc}
\hline
Dataset & \multicolumn{1}{l}{Varying parameter} & Weighted vs. Unweighted 
& WSB-50 vs. Unbinned & WSB-75 vs. Unbinned\\ \hline
10-taxon  & \multicolumn{1}{l}{ILS level} & 0.96 & 0.96 & 0.96 \\ 
15-taxon & \multicolumn{1}{l}{\# of genes, seq length} & 0.96 & 0.96 & \textbf{0.04} \\ 
Avian  & \multicolumn{1}{l}{sequence length} & 0.96 & \textbf{\textless0.0001} & n.a. \\ 
Avian  & \multicolumn{1}{l}{ILS level} & 0.96 & \textbf{\textless0.0001}  & n.a. \\ 
Avian  & \multicolumn{1}{l}{\# of genes} & 0.91 & \textbf{\textless0.0001}  & n.a. \\ 
Mammalian  & \multicolumn{1}{l}{\# of genes, seq length} & 0.96 & n.a. & \textbf{\textless0.0001}   \\ 
Mammalian  & \multicolumn{1}{l}{ILS level} & 0.96 & n.a.  &  \textbf{0.0003}\\ 
\end{tabular}
\end{small}
\caption{{\bf Statistical significance 
test results for choice of binning method on
MP-EST.} We performed ANOVA to test the significance of the choice of
methods (unbinned, weighted binned, unweighted binned, WSB-50: weighted 
statistical binning using 50\% bootstrap support threshold and WSB-75: weighted binning using 75\% bootstrap support threshold). For weighted vs. unweighted, we compared 50\% bootstrap support threshold for avian, 75\% for mammalian, and both 50\% and 75\% for 15- and 10-taxon datasets. All $p$-values are corrected for multiple hypothesis testing using the FDR correction ($n=16$). ``n.a.'' stands for ``not available''.}
\label{table-test-mpest}
\end{table}

\begin{table}[h]
\begin{small}
\begin{tabular}{llccc}

\hline
Dataset & \multicolumn{1}{l}{Varying parameter} & Weighted vs Unweighted & WSB-50 vs Unbinned & WSB-75 vs Unbinned\\ \hline
10-taxon  & \multicolumn{1}{l}{ILS level} & 1 & 1 & 0.91 \\ 
15-taxon & \multicolumn{1}{l}{\# of genes, seq length} & 0.91 & 0.57 & \textbf{0.008} \\ 
Avian & \multicolumn{1}{l}{sequence length} & 0.91 & \textbf{\textless0.0001} & n.a. \\ 
Avian  & \multicolumn{1}{l}{sequence length}  & 1 & n.a. & 0.57 \\ 
Mammalian  & \multicolumn{1}{l}{ILS level}  & 0.57 & n.a. & \textbf{0.0009} \\ 
Mammalian  & \multicolumn{1}{l}{\# of genes}  & 0.91 & n.a. & \textbf{\textless0.0001}  \\

\end{tabular}
\end{small}
\caption{{\bf Statistical significance 
test results for choice of binning method on ASTRAL.} We performed ANOVA to test the significance of the choice of
methods (unbinned, weighted binned, unweighted binned, WSB-50: weighted 
statistical binning using 50\% bootstrap support threshold and WSB-75: weighted binning using 75\% bootstrap support threshold). For weighted vs. unweighted, we compared 50\% bootstrap support threshold for avian, 75\% for mammalian, and both 50\% and 75\% for 15- and 10-taxon datasets. All $p$-values are corrected for multiple hypothesis testing using the FDR correction ($n=14$). ``n.a.'' stands for ``not available''.}
\label{table-test-astral}
\end{table}

The largest improvements are for the shortest gene sequences,
where error is reduced from 23\% to 14\% using  
MP-EST and from 19\% to 13\% using ASTRAL.
The difference between binned and unbinned analyses is lower
for 1000bp sequences, 
and there are no noteworthy differences for
1500bp sequences (sequence length has
a statistically significant impact; see 
Tables~\ref{table-test-interaction-mpest} and \ref{table-test-interaction-astral}).
When the number of genes is changed (see Fig.~\ref{fig-fn-avian}(c)),
the impact of binning on MP-EST ranges from neutral to
highly positive, and the largest improvements are for
datasets with large numbers of genes
(impact of the number of genes is statistically significant; 
see Table~\ref{table-test-interaction-mpest}).
The impact of binning is also significantly impacted by ILS levels 
(see Table \ref{table-test-interaction-mpest}), 
with the largest improvements obtained for lower levels of ILS.
In general, binning helps both ASTRAL and MP-EST, but MP-EST tends to be 
helped more than ASTRAL. For example, 
with 500bp genes, the error for MP-EST is reduced from
19\% to 10\% using binning, but error of ASTRAL is reduced 
from 15\% to 9\%. 

\begin{table}[h]
\begin{small}
\begin{tabular}{llccc}

\hline
Dataset & \multicolumn{1}{l}{Interaction variable} & Weighted vs Unweighted & WSB-50 vs Unbinned & WSB-75 vs Unbinned\\ \hline
10-taxon  & \multicolumn{1}{l}{ILS level} & 0.99 & 0.99 & 0.49 \\ 
15-taxon & \multicolumn{1}{l}{\# of genes, seq length} & 0.99 \& 0.99 & 0.59 \& 0.99 & 0.24 \& 0.17 \\ 
Avian  & \multicolumn{1}{l}{sequence length} & 0.99 & \textbf{\textless0.0001} & n.a. \\ 
Avian  & \multicolumn{1}{l}{ILS level} & 0.99 & \textbf{\textless0.0001}  & n.a. \\ 
Avian  & \multicolumn{1}{l}{\# of genes} & 0.99 & \textbf{\textless0.0001}  & n.a. \\ 
Mammalian  & \multicolumn{1}{l}{\# of genes, seq length} & 0.99 \& 0.99 & n.a. & 0.99 \& 0.38   \\ 
Mammalian  & \multicolumn{1}{l}{ILS level} & 0.15 & n.a. &  0.15\\ 
\end{tabular}
\end{small}
\caption{{\bf Statistical significance test results for 
interaction effects (binning and simulation parameter) on MP-EST.}
 We performed ANOVA to test the significance of whether there is an interaction between the choice of the method (unbinned, weighted binned, unweighted binned, WSB-50: weighted statistical binning using 50\% bootstrap support threshold and WSB-75: weighted statistical binning using 75\% bootstrap support threshold) and the variable changed in each dataset. For weighted vs. unweighted, we compared 50\% bootstrap support threshold for avian, 75\% for mammalian, and both 50\% and 75\% for 15- and 10-taxon datasets. All $p$-values are corrected for multiple hypothesis testing using the FDR correction ($n=21$). ``n.a.'' stands for ``not available''.}
\label{table-test-interaction-mpest}
\end{table}

\begin{table}[h]
\begin{small}
\begin{tabular}{llccc}

\hline
Dataset & \multicolumn{1}{l}{Interaction variable} & Weighted vs Unweighted & WSB-50 vs Unbinned & WSB-75 vs Unbinned\\ \hline
10-taxon  & \multicolumn{1}{l}{ILS level} & 0.99 & 1 & 0.99 \\ 
15-taxon & \multicolumn{1}{l}{\# of genes, seq length} & 0.99 \& 0.99 & 0.99 \& 0.99 & 0.29 \& \textbf{0.02} \\ 
Avian & \multicolumn{1}{l}{sequence length} & 0.99 & \textbf{\textless0.0001} & n.a. \\ 
Mammalian  & \multicolumn{1}{l}{sequence length} & 0.99 & n.a. & 0.29 \\ 
Mammalian  & \multicolumn{1}{l}{ILS level} & 0.99 & n.a. & 0.29 \\ 
Mammalian  & \multicolumn{1}{l}{\# of genes} & 0.99 & n.a. & 0.99  \\ 

\end{tabular}
\end{small}
\caption{{\bf Statistical significance test results for 
interaction effects (binning and simulation parameter) on ASTRAL.}
 We performed ANOVA to test the significance of whether there is an interaction between the choice of the method (unbinned, weighted binned, unweighted binned, WSB-50: weighted statistical binning using 50\% bootstrap support threshold and WSB-75: weighted statistical binning using 75\% bootstrap support threshold) and the variable changed in each dataset. For weighted vs. unweighted, we compared 50\% bootstrap support threshold for avian, 75\% for mammalian, and both 50\% and 75\% for 15- and 10-taxon datasets. All $p$-values are corrected for multiple hypothesis testing using the FDR correction ($n = 17$). ``n.a.'' stands for ``not available''.}
\label{table-test-interaction-astral}
\end{table}

Fig.~\ref{fig5} shows the impact of binning on
species tree branch length estimation error on the
biologically-based  simulations using MP-EST;
Fig.~\ref{fig5}(a) shows results on 1000 genes under
default (1X) ILS levels and varying
gene sequence length, and
Fig.~\ref{fig5}(b) shows results
on 1000 genes of 500bp with varying ILS levels. 
Branch length estimation accuracy is reported using the
ratio of the estimated branch length to the true branch
length, for those true branches 
recovered by the method.
Thus, values equal to 1 indicate perfect
accuracy, values below 1 indicate under-estimation of
branch lengths (and hence over-estimation of ILS),
and values above 1 indicate over-estimation of branch
lengths (and hence under-estimation of ILS).

Both types of binning (weighted and unweighted)
produce nearly identical results with respect to  species tree 
branch length estimation (with a slight advantage for weighted analyses).
Unbinned analyses substantially
under-estimate branch lengths, but
as the sequence length increases, the branch
length estimations produced
by unbinned analyses improve, so that they are
more  accurate with 1500bp
markers. 
The most accurate  species tree 
branch length estimation is obtained using true gene trees.
Using binning (either type) improves branch 
length estimation from estimated gene trees, and the improvement is
very large for the shorter sequences (Fig.~\ref{fig5}(a)). 
When levels of ILS are changed,
weighted and unweighted binning are again close (with a slight advantage for weighted),
and show little change in
branch length estimation with changes in ILS
levels; however, unbinned analyses substantially
under-estimate branch lengths for the lowest ILS
model condition, and then become more accurate (although
still under-estimate) with increases in the ILS level.
Hence, 
the biggest improvement obtained by binning
is for the lowest ILS (2X branch lengths), and there is
less improvement for the highest ILS level (0.5X).
The likely explanation for
this trend 
is that MP-EST interprets all
discord as due to ILS, and produces
a model tree (with branch lengths) that it considers
most likely to generate the observed discordance. Hence,  MP-EST 
tree branch lengths will be closer to
the correct lengths when the ILS level is very high.

\subsubsection*{Bootstrap support on avian simulated datasets}

We explore bootstrap support of trees estimated
on simulated avian datasets, as follows.
We assign relative quality to each edge in an
estimated tree, taking bootstrap support into account.
The highest quality edges are the true positive branches
with the highest bootstrap support, and the
lowest quality edges are the false positive branches
with the highest bootstrap support, and all other edges
fall in between.
We order all the edges by their quality, so that
the true positive branches come first 
(with the high support branches before low support
branches), followed by
the false positive branches 
(with the low support branches 
before the high support branches).
Given this ordering, we create figures in which  the
x-axis indicates the edge quality  (from very high to very low,
as you move from left to right), and the y-axis indicates the 
fraction of the edges having at least the quality indicated by
the x-axis. Thus, the higher the curve, the better the overall
quality  of
the species tree. 

Fig.~\ref{fig:tp-fp-avian-bpILS}  shows
results on 1000 avian 
genes 
under default ILS and with varying
sequence length, and also with 
1000 genes of 500bp with varying ILS levels.
Both types of binning are nearly
identical in terms of their impact on bootstrap
support, and both improve bootstrap support; in particular,
using binning increases the number of highly 
supported true positive branches and decreases the number of
highly supported false positive branches.
However, the sequence length
modulates the impact of binning on
bootstrap support, so that the
largest impact is for the shortest sequences
(250bp)
and there is no discernible impact for the longest
sequences (1500bp).  
ILS levels also impact how binning
affects the bootstrap support, so that
the biggest improvement in bootstrap support 
is obtained for the lowest ILS level (2X branch lengths).
 The number of genes also impacts the bootstrap support (supporting
information S2 Fig.). 
 so
that the biggest improvement in bootstrap support
is obtained for the largest number of genes (2000) 
(and there is little to no difference between 
binned and unbinned analyses on 50 or 100 genes);
furthermore, weighted and unweighted binning
produce very similar bootstrap support values.

\subsubsection*{Comparisons to concatenation on avian simulated datasets}
\begin{figure}[!ht]
 \begin{center}
\includegraphics[width=0.8\textwidth]{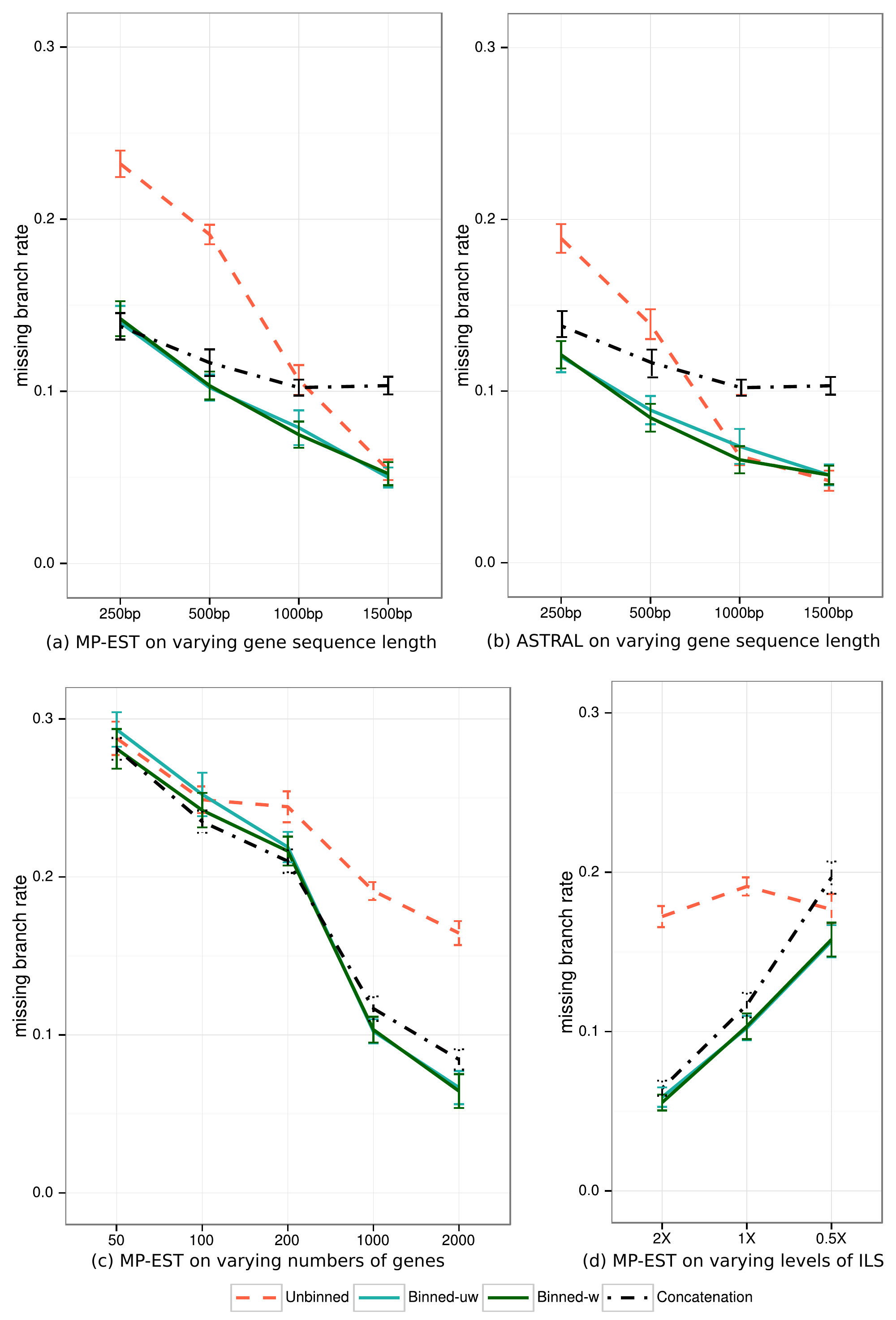}
 \end{center}
 \caption{{\bf Species tree estimation error (FN) for MP-EST and ASTRAL with MLBS on avian
simulated datasets}.  (a) MP-EST on 1000 genes with
varying gene sequence length (controlling gene tree error) and
with 1X ILS.
(b) ASTRAL on the exact same conditions,
(c) MP-EST on varying numbers of genes with fixed default level of ILS (1X level) and 500bp sequence length,  and (d) MP-EST on varying levels of ILS and 1000 genes of length 500bp.
We show results for 20 replicates everywhere, except for 2000 genes that are based on 10 replicates. Binning was performed using 50\% bootstrap support threshold. We estimated the supergene trees, and performed concatenation using RAxML with unpartitioned analyses.}
 \label{fig-fn-avian}
 \end{figure}

\begin{figure}[!ht]
\begin{center}
\includegraphics[width=0.92\textwidth]{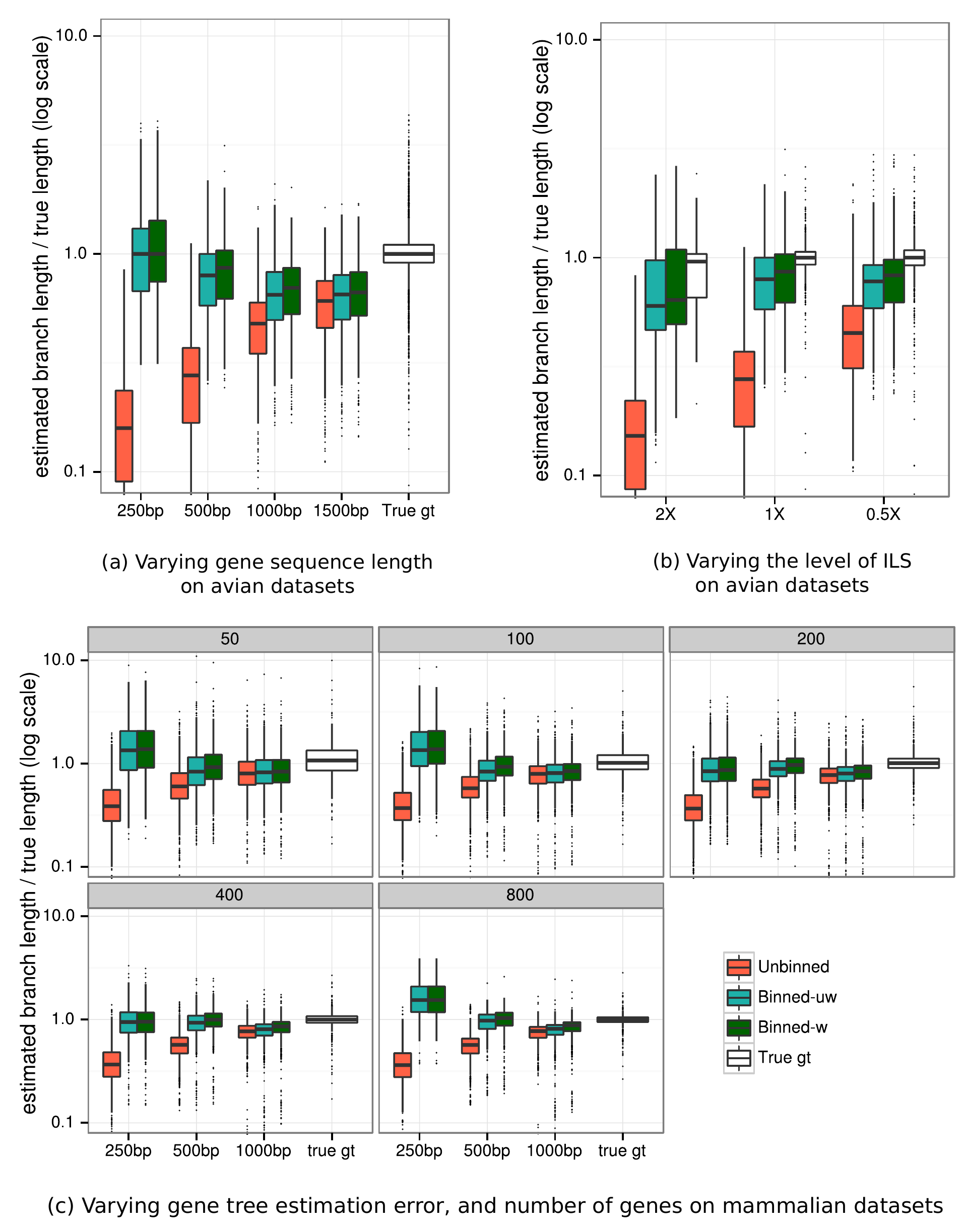}
\end{center}
\caption{{\bf Effect of binning on the branch lengths (in coalescent units) estimated by MP-EST using MLBS on the avian and mammalian simulated datasets}. We show the species tree branch length error (the ratio of estimated branch length to true branch length for branches of the true tree that appear in the estimated tree; 1 indicates correct estimation). Results are shown for (a) 1000 avian genes of 1X ILS level with varying gene sequence length, (b) 1000 avian genes of 500bp and with varying levels of ILS, and (c) varying number of mammalian genes and varying sequence length (250bp, 500bp, and 1000bp) with 1X  ILS level. Results are shown for 20 replicates. We used 50\% and 75\% bootstrap support threshold for binning on avian and mammalian datasets, respectively,  and estimated the supergene trees using RAxML with unpartitioned analyses.
}
\label{fig-mammal-bl}
\label{fig5}
\end{figure}

\begin{figure}[h]
\begin{center}
\includegraphics[width=0.97\textwidth]{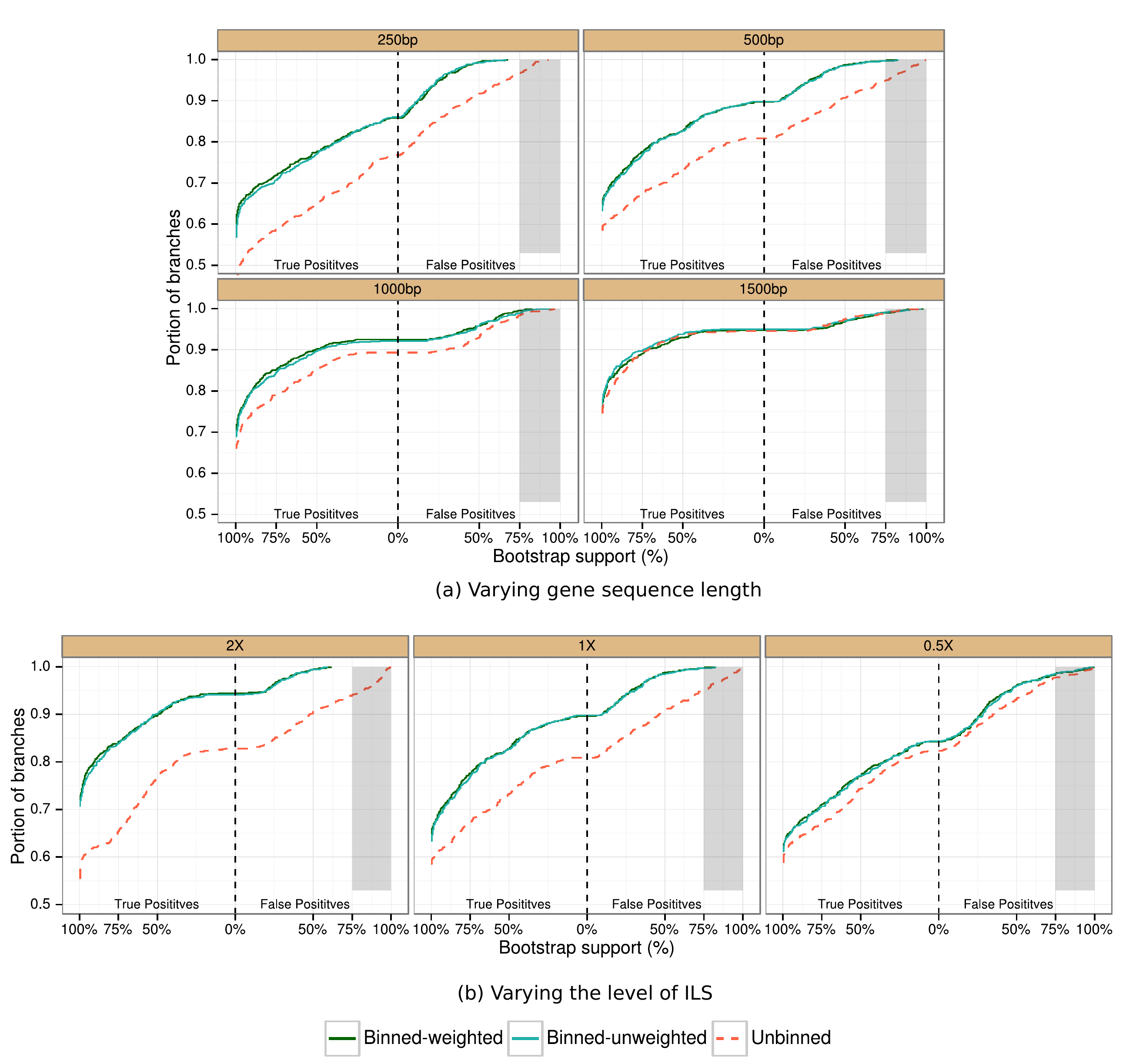}
\end{center}
\caption{{\bf Cumulative distribution of the 
bootstrap support values (obtained using MLBS) of 
true positive (TP) and false positive (FP) edges estimated 
by binned and unbinned MP-EST on avian datasets}. In (a) we fix the number
of genes to 1000, use default ILS levels, and vary
sequence length to control 
gene tree estimation error, and in (b) we study 1000 genes with 500bp sequence length, and vary ILS levels. 
To produce the graph, we order the branches in the estimated species tree by their quality,
so that the true positives with high support come first, followed by lower support true positives,
then by false positives with low support, and finally by false positives with high support.
The false positive branches with support above 75\% are the most troublesome, and
the highly supported false positives are indicated by the grey area. When the curve for a method lies above the
curve for another method, then the first method has better bootstrap support. We used 50\% bootstrap support threshold for binning, and estimated the supergene trees using RAxML with unpartitioned analyses.
} \label{fig:tp-fp-avian-bpILS}
\end{figure}

On the shortest 250bp sequences, concatenation matches the accuracy
of weighted and unweighted binned MP-EST methods (Fig.~\ref{fig-fn-avian}(a)) 
and is slightly less accurate than both binned ASTRAL trees (Fig.~\ref{fig-fn-avian}(b)).
As sequence length increases, both types of binning using either ASTRAL or MP-EST
become more accurate than concatenation.
Unbinned analyses are less accurate than 
concatenation for shorter sequences and more accurate for longer
sequences (the transition point depends on whether ASTRAL or MP-EST is used).
Both binned analyses 
are more accurate than concatenation and
unbinned analyses at all ILS levels
(Fig.~\ref{fig-fn-avian}(d)).
Thus, compared to concatenation, binned analyses
have their largest advantage on longer gene sequences,
higher ILS levels, and higher number of genes.

\subsubsection*{Results on mammalian datasets}
\begin{figure}[!ht]
\begin{center}
\includegraphics[width=0.93\textwidth]{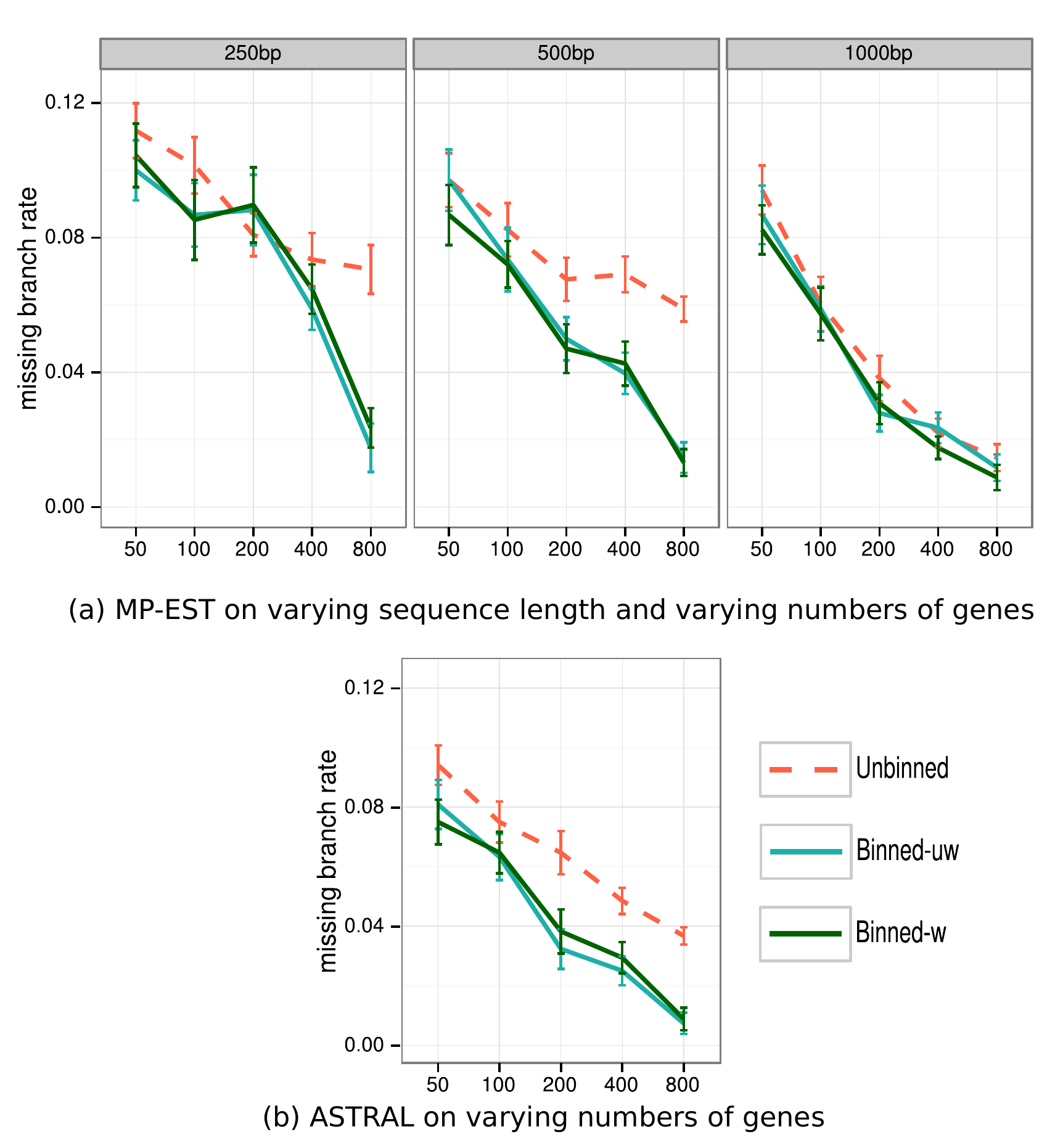}
\end{center}
\caption{{\bf Species tree estimation error for MP-EST and ASTRAL using MLBS on
mammalian simulated datasets}. We show average FN rate over 20
replicates. (a) Results for MP-EST. We varied the number of genes (50, 100, 200, 400 and 800) and
sequence length (250bp (43\% BS), 500bp (63\% BS) and 1000bp (79\% BS)) with default
amount of ILS (1X level). (b) ASTRAL on varying numbers of genes with fixed 1X ILS level and 500bp sequence length. We used 50\% and 75\% bootstrap support threshold for binning on avian and mammalian datasets, respectively,  and estimated the supergene trees using RAxML with unpartitioned analyses. } \label{fig:mammal-main-mpest}
\end{figure}

Results on simulated mammalian datasets are similar
to analyses of avian datasets.
 In nearly every condition,
both weighted and unweighted binning show very
similar results (see Fig.~\ref{fig:mammal-main-mpest}) and 
have no statistically significant differences using either ASTRAL or MP-EST 
(see Tables~\ref{table-test-mpest} and \ref{table-test-astral}).
As before, we evaluated the impact of statistical binning
on gene tree estimation error under the 1X (default ILS) model
condition with varying sequence lengths (S1 Table), and
observed that binning substantially reduces gene tree estimation
error for short sequences (250bp and 500bp) but had
little impact on longer sequences (1000bp). 
Binning improves gene tree distributions, 
generally
with very large
improvements,  and the improvements
decrease with the 
sequence length and ILS level (S3 Fig.). 
Binning also improves species tree
topology estimation 
(Fig.~\ref{fig:mammal-main-mpest} and Tables~\ref{table-test-mpest} and \ref{table-test-astral}).
The impact appears to depend on the
sequence length (binning seems more beneficial for shorter sequences
and neutral for longer sequences)
and number of genes (binning can dramatically improve
species tree topologies
given a large number of genes, but can be neutral or
even detrimental for a small number of genes), 
and the choice of summary method (binning helps
both ASTRAL and MP-EST, but helps MP-EST more).
ILS level also seems to impact relative accuracy
(Tables~\ref{table-test-interaction-mpest} and \ref{table-test-interaction-astral}), so that binning seems most
helpful for low ILS levels, and less helpful for high ILS
levels (S4 Fig.). 
However, the effects of number of genes, sequence length, 
and the ILS level were not 
statistically significant for this dataset 
(Tables~\ref{table-test-interaction-mpest} and \ref{table-test-interaction-astral}).

As observed in the avian simulations, 
unbinned analyses substantially under-estimate species
tree branch lengths (Fig.~\ref{fig5}(c) and S5 Fig.). 
Both weighted and unweighted
binning produce nearly identical branch lengths
for all sequence lengths,  number of genes, and ILS
levels, 
and both types of binning come closer to the
true branch lengths
than unbinned analyses.
Finally, both weighted and unweighted binning produce
nearly identical 
species tree branch support values, where both match or
improve
unbinned analyses for 
all tested numbers of genes, sequence lengths, and ILS levels
(S6 and S7 Figs.).  
However, 
improvements increase with the number of genes and decrease with
the sequence length and ILS level.

\subsubsection*{Impact of support threshold $B$ on avian and
mammalian simulated datasets}
\begin{figure}[!ht]
\begin{center}
\includegraphics[width=0.6\textwidth]{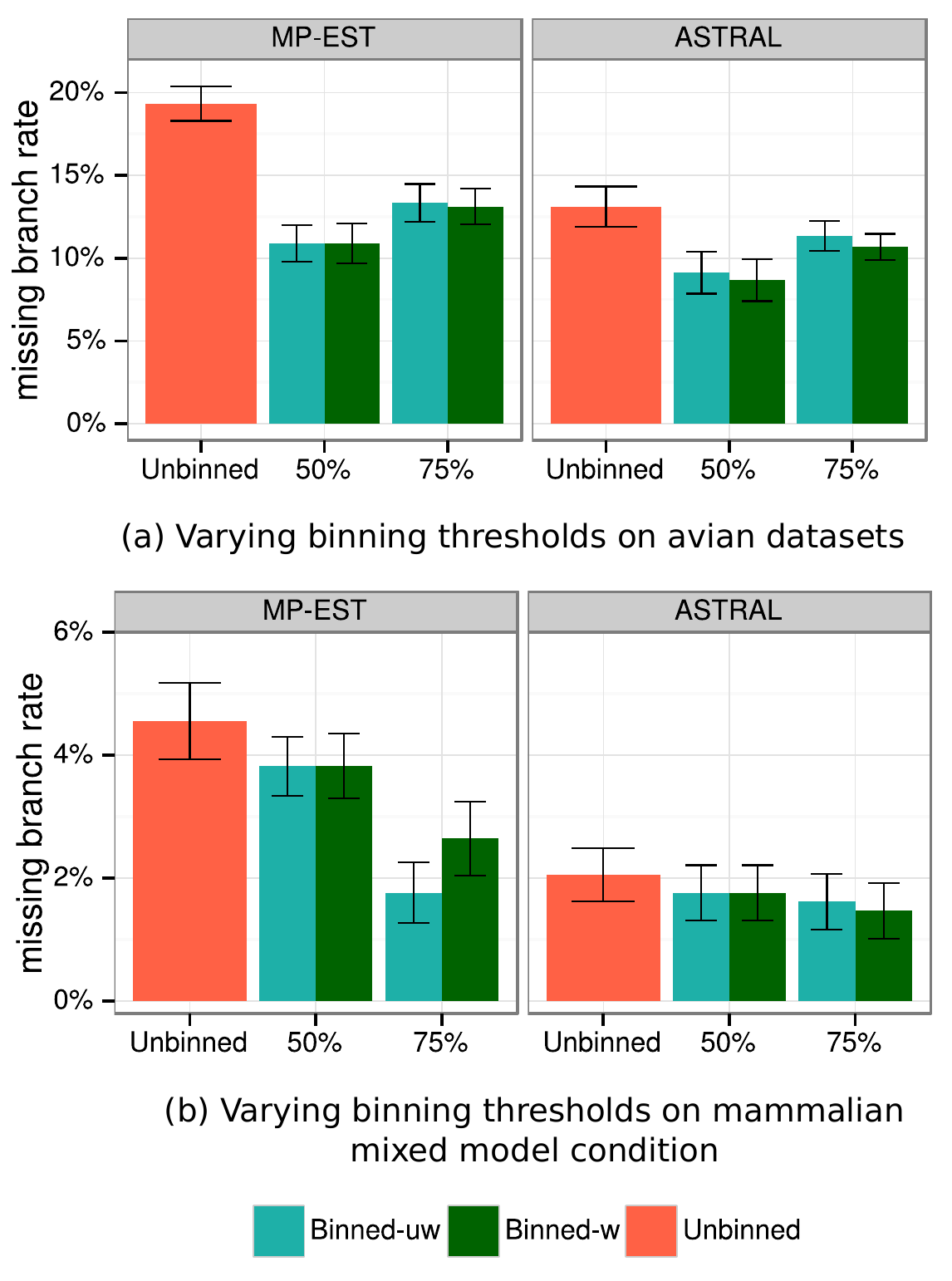}
\end{center}
\caption{{\bf Species tree estimation error for MP-EST and ASTRAL using MLBS on
avian and mammalian simulated datasets with two support thresholds ($B$)}. 
We show average FN rate for unbinned, and wighted and unweighted binned analyses
with both $B=50\%$ and $B=75\%$.
Results are shown for (a)  the avian dataset with 10 replicates of 
1000 genes of length 500bp and 1X ILS level, and
(b) the mammalian dataset with 20 replicates of 
400 mixed genes (200 genes with 500bp and 200 genes with 1000bp) with 1X ILS level.
} \label{fig-thresholds}
\end{figure}

In addition to varying model conditions, 
we use a single avian and a single mammalian model condition 
to study the 
impact of the support threshold $B$ on binning (Fig.~\ref{fig-thresholds}).
We use a mixed model condition with
200 genes of 500bp and 200 genes of 1000bp
for the mammalian dataset, 
and a model condition with 1000 genes of 500bp for the avian dataset
(both with default 1X ILS level).

On the avian dataset, binning is always beneficial, 
but the impact is larger with $B=50\%$ compared to $B=75\%$ (Fig.~\ref{fig-thresholds}(a)). 
For example, unbinned MP-EST has 19\% error,
and using  $B=50\%$ reduces the error to
11\%, and using $B=75\%$ reduces  the error 
to 13\%.  

On the mammalian mixed data, 
binning is beneficial in all cases (see Fig.~\ref{fig-thresholds}(b));
however, the extent of the 
impact depends substantially on both the threshold and the summary method. 
ASTRAL has high accuracy even without binning, 
and binning with either threshold has only a small impact on its accuracy.  
When MP-EST is used, binning with $B= 50\%$ 
leads to relatively small improvements in accuracy,
whereas $B= 75\%$ results in much larger improvements. 
Thus, the choice of the threshold can have an impact, but for the two model
conditions we studied here both choices of the threshold are beneficial.

\subsubsection*{Effects of binning on gene tree and species tree error for 15-taxon datasets}

We explored the impact of statistical binning
on gene tree estimation error using two sequence lengths and two
values for $B$, the
bootstrap support threshold parameter (S1 Table).
For the shorter sequence lengths (100bp), binning increases
gene tree estimation error (from 77\% to 80\% when $B=50\%$, and
from 77\% to 86\% when $B=75\%$). For the longer sequence lengths (1000bp),
binning with $B=50\%$ has no impact on gene tree estimation error,
but using $B=75\%$ increases error from 36\% to 40\%.  Thus, 
statistical binning increases gene tree estimation error for
these very high ILS
15-taxon datasets, but the amount of the increase depended on the
parameter $B$ (with larger increases for $B=75\%$ and small
increases for $B=50\%$)  and sequence length (where the
impact on gene tree estimation error is much reduced for
the 1000bp alignments).

\begin{figure}[!ht]
\begin{center}
\includegraphics[width=0.90\textwidth]{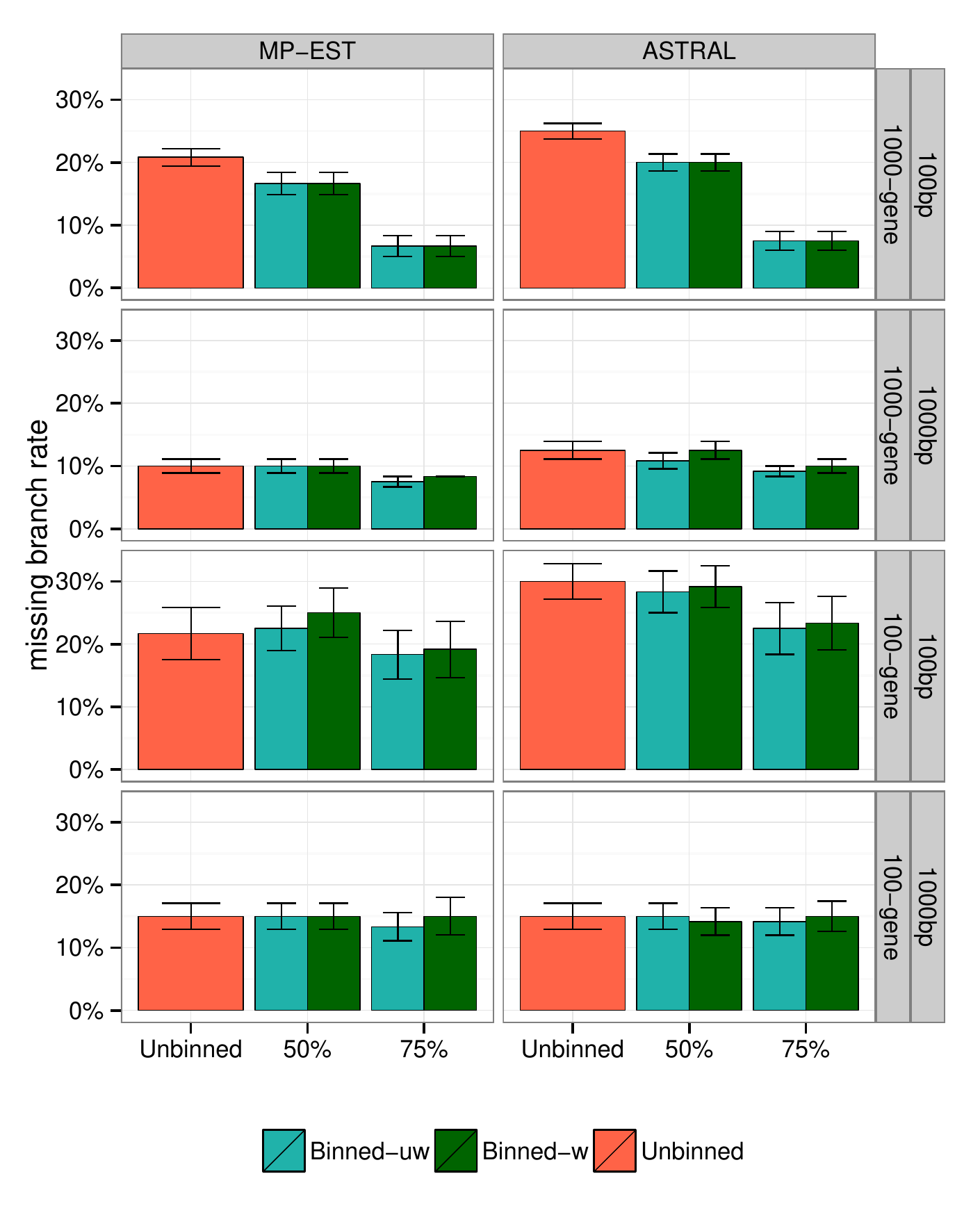}
\end{center}
\caption{{\bf Species tree estimation error for MP-EST and ASTRAL with MLBS on
15-taxon simulated datasets}. We show average FN rate over 10
replicates. We varied the number of genes (100 and 1000) and sequence length (100bp and 1000bp). We used 50\% and 75\% bootstrap support thresholds for binning, and estimated the supergene trees using RAxML with fully partitioned analyses.} \label{fig:15-taxon-mpest-astral}
\end{figure}

Fig.~\ref{fig:15-taxon-mpest-astral} shows the impact of weighted and unweighted 
statistical binning on species tree accuracy for the 15-taxon dataset.
We apply statistical binning with two support thresholds (50\% and 75\%), 
and we use both MP-EST and ASTRAL as the summary method.
In all cases, weighted and unweighted binning have similar accuracy,
with 
no statistically significant differences (Tables \ref{table-test-mpest} and \ref{table-test-astral}).
The relative
accuracy  of unbinned and binned analyses depends on the support threshold, 
so that with $B=50\%$, there are no statistically significant 
differences, but with $B=75\%$,
binning significantly improves accuracy ($p=0.04$ for 
MP-EST and $p=0.008$ for ASTRAL; Tables \ref{table-test-mpest} and \ref{table-test-astral}).
The extent of the improvements seems larger 
for more genes and smaller alignments, but the impact
of these factors are not statistically significant for MP-EST ($p=0.24$ and $p=0.17$ respectively)
and only impact of sequence length was significant for ASTRAL ($p=0.02$ ; Tables \ref{table-test-interaction-mpest} and \ref{table-test-interaction-astral}). 
The biggest gains are obtained when the 75\% threshold is used
with 1000 genes of 100bp, where binning reduces the error of MP-EST 
from 21\% to only 7\%. 
Thus, the choice of the threshold can matter, and on this dataset, the 
effects of binning can range from neutral to highly beneficial, 
depending on the threshold used, number of genes, and gene sequence length. 

\subsubsection*{Effects of binning on species tree error for 10-taxon datasets}
\begin{figure}[!ht]
\begin{center}
\includegraphics[width=0.93\textwidth]{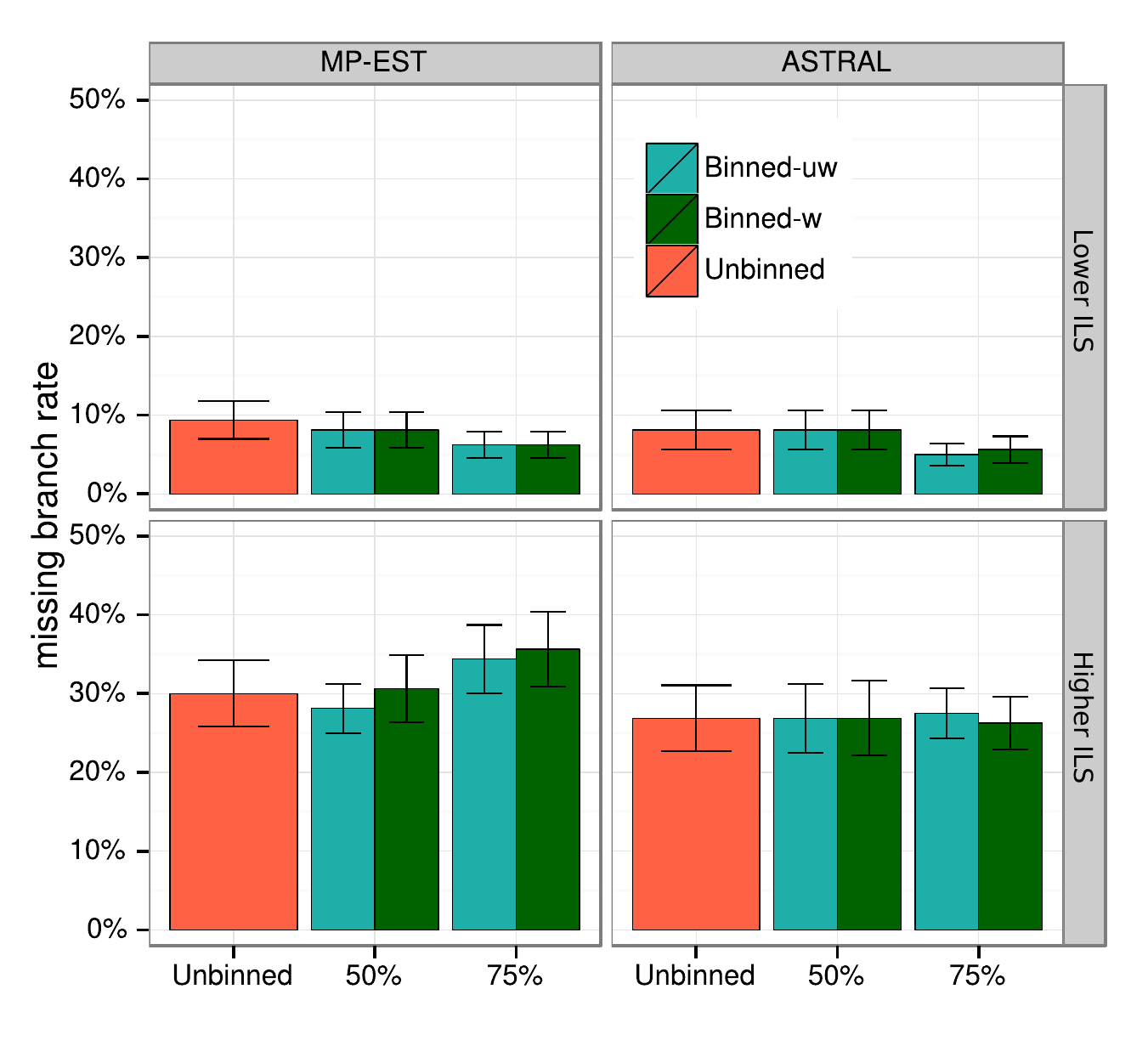}
\end{center}
\caption{{\bf Species tree estimation error for MP-EST and ASTRAL with MLBS on
10-taxon simulated datasets}. We show average FN rate over 20
replicates. We varied the amount of ILS and fixed the number of genes to 200 and gene sequence length to 100bp. We used 50\% and 75\% bootstrap support thresholds for binning, and estimated the supergene trees using RAxML with fully partitioned analyses.} \label{fig:10-taxon-mpest-astral}
\end{figure}

Fig.~\ref{fig:10-taxon-mpest-astral} shows the impact of binning on species 
tree accuracy on the 10-taxon 
datasets with two choices of the threshold $B$ for the statistical binning pipeline
($B=50\%$ and $75\%$), two choices of the summary method (MP-EST and ASTRAL), 
and two levels of ILS (high and very high).
No statistically significant differences are observed on these data between
weighted and unweighted binning, or between weighted binning and unbinned analyses
(see Tables \ref{table-test-mpest} and \ref{table-test-astral}); nevertheless, some patterns can be observed 
in terms of the average error (Fig.~\ref{fig:10-taxon-mpest-astral}).
Both weighted and unweighted statistical binning 
are close to neutral (regardless of the choice of method or level of ILS)
when applied with a 50\% threshold. 
When the 75\% threshold is used, the impact of binning
depends on the level of ILS:
binning improves accuracy with low ILS levels
and 
reduces accuracy with high ILS levels, especially
when MP-EST is used, but 
these differences are not statistically 
significant (Tables \ref{table-test-mpest} and \ref{table-test-astral}).

\subsection*{Analysis of biological datasets}
We compared 
weighted and unweighted binning of MP-EST and ASTRAL
on MLBS gene trees
on the avian and mammalian  biological datasets studied in
\cite{MirarabStatBinning}. 

\begin{figure}[!ht]
 \begin{center}
\includegraphics[width=0.98\textwidth]{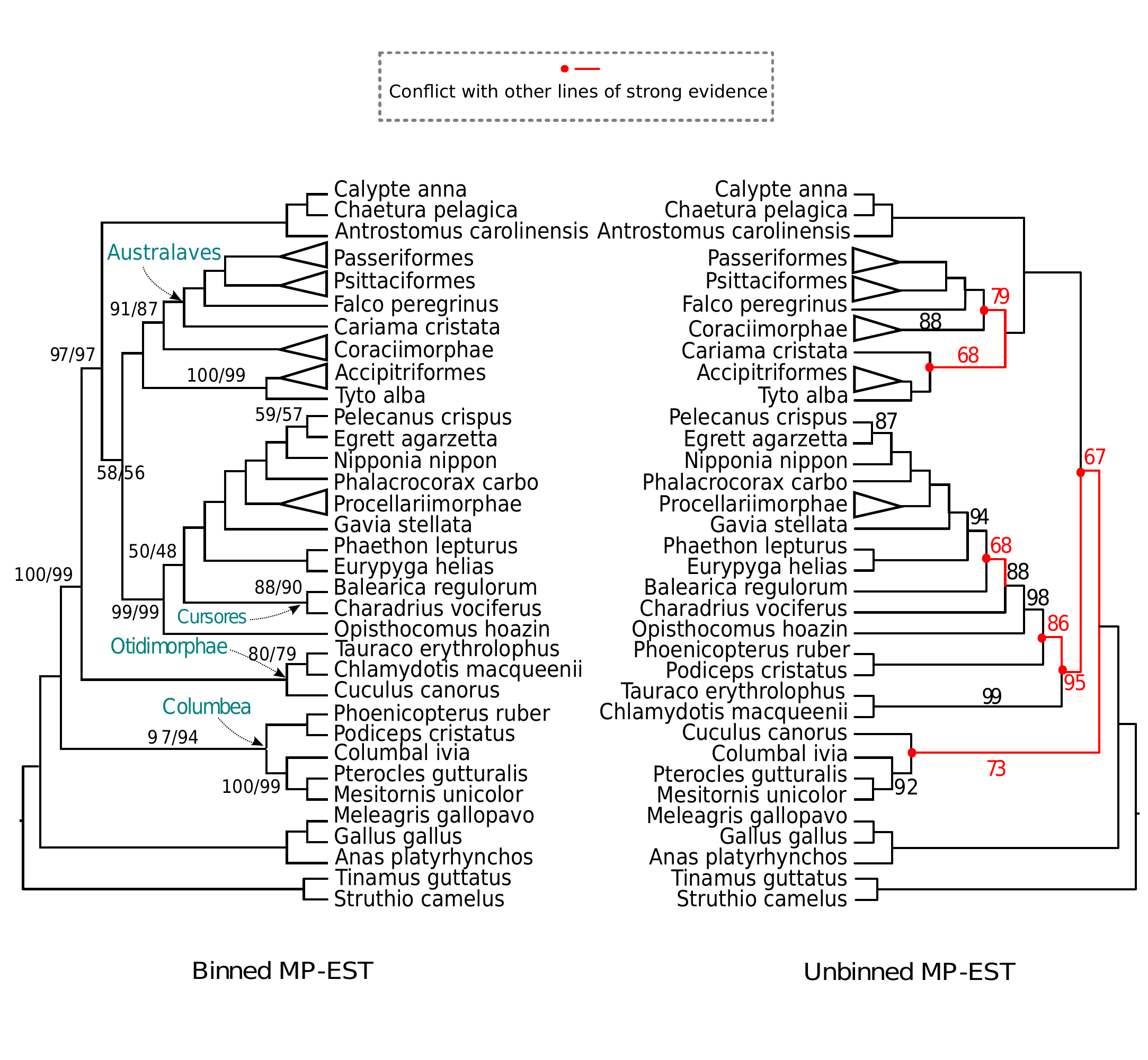}
 \end{center}
 \caption{
{\bf Trees computed on the avian biological dataset using MP-EST on MLBS gene
trees. }
We show results
with weighted
and unweighted binning (left), and unbinned analyses (right).  We used 50\% bootstrap support threshold for binning. Supergene trees were estimated using fully partitioned analyses.
MP-EST with weighted and unweighted binning returned the same tree. 
The branches on the binned MP-EST tree are labeled with two support values side by side: 
the first is for unweighted binning and the second is for weighted binning; branches 
without designation have 100\% support.
Branches in red indicate contradictions to 
known subgroups.}
 \label{fig:avian-bio}
 \end{figure}

Results for MP-EST on these datasets showed the following
trends. 
First, 
for the avian dataset, 
there are
no topological differences
between MP-EST trees estimated using weighted or unweighted statistical
binning, 
and extremely small
differences in branch support (less than 3\%; see Fig.~\ref{fig:avian-bio}). 
Thus, although \cite{JarvisScience2014} only 
explored  unweighted statistical
binning with MP-EST, 
the main conclusions they drew about the evolutionary history of 
modern birds are also found in the weighted statistical binning
analysis using MP-EST.
The unbinned MP-EST analysis violates
several subgroups established in the avian phylogenomics 
project and other studies (indicated in red in Fig.~\ref{fig:avian-bio}), but the
binned MP-EST analyses do not violate any of these subgroups.
Of these violated subgroups, the failure of the unbinned 
MP-EST analysis to
recover Australaves is the most significant, since it has been
recovered in many prior analyses \cite{SuhNatComm2011,WangMBE2012,KimballMPE2013,McCormack2013}.
On the mammalian dataset, weighted and unweighted MP-EST
again produce the same exact tree, with small differences in 
support (less than 3\%; see S11 Fig.).  
The unbinned MP-EST tree, however, 
has one topological difference (the position of treeshrews;
compare S10 and S11 Figs.). 
with binned analyses, as discussed 
in \cite{MirarabStatBinning}.

Results for ASTRAL on the biological datasets show generally similar trends.
Unbinned ASTRAL on the avian dataset 
(S8 Fig.) 
recovers Australaves 
and hence is more in line with 
the prior literature than unbinned MP-EST;
however, just like unbinned MP-EST, the unbinned ASTRAL  does not recover some key clades
recovered by concatenation and other analyses reported in \cite{JarvisScience2014}.
Using weighted and unweighted statistical binning with 
ASTRAL on the avian dataset
produces almost identical results, 
and are also almost identical to the binned MP-EST tree
(the only change is 
the position of hoatzin which has low support in all trees; see supporting
information S9 Fig.). 
On the mammalian dataset,
trees produced by binned ASTRAL analyses 
with weighted or unweighted binning pipelines are 
topologically identical to each other, and to the tree produced 
by the unbinned analysis,
and have rather small differences
in bootstrap support (see S12 Fig.).  
Binned and unbinned ASTRAL analyses and binned MP-EST analyses
all put treeshrews as sister to Glires, 
while unbinned MP-EST puts them as sister to primates.
The placement of treeshrews is of substantial debate, and
so the differential placement is of considerable interest
in mammalian systematics. 

Overall, results on the two biological datasets show that weighted
and unweighted statistical binning analyses
produced identical species trees 
and nearly identical branch support values; furthermore, these binned
analyses were more congruent with established
subgroups than unbinned analyses. 


\subsection*{Summary of observations}
Across all our analyses, results for both ASTRAL and MP-EST are
very similar with respect to how they responded to statistical binning.
Weighted statistical binning 
produces nearly identical results
to unweighted statistical binning on the
biologically-based
simulated datasets, 
and topologically identical results (with
very similar bootstrap support values) on the biological
datasets we explored in this study, and so this study 
generally supports the conclusions about statistical binning
in \cite{MirarabStatBinning}. In addition, 
because weighted and unweighted statistical binning
produce topologically identical trees on the avian dataset, 
this study supports the
findings about the avian phylogeny reported
in \cite{JarvisScience2014}.  
The fact that weighted and unweighted binning typically produced similar results
is not surprising, since the unweighted binning technique strives to 
create ``balanced'' bins as much as possible, and largely
achieves this on the datasets we explored.
Furthermore,  if the bins produced by statistical binning have {\em exactly} the same size,
then pipelines based on weighted and unweighted statistical binning will produce the same species tree,
since the distributions of gene trees they produce will be identical.
Since the bin sizes produced using our heuristic for balanced
minimum vertex coloring are close to balanced, 
this explains
why we observed very small  differences between weighted and unweighted statistical binning
in these analyses.

Under most of the model conditions we studied, 
both weighted and unweighted statistical binning
improved the estimation of gene tree topologies, 
gene tree distributions, 
species tree topologies and branch lengths, 
and bootstrap support (so that
statistical binning 
increases bootstrap support for true positive edges,
and reduces the number of highly supported false positives),
compared to unbinned analyses.
These improvements are largest 
when gene sequence alignments have low
phylogenetic signal, the gene trees
exhibit  at most moderately large
ILS levels, or there are many genes.


The impact of statistical binning on the 15-taxon
datasets is somewhat different than for the biologically-based
simulations. Gene tree estimation accuracy is reduced
for both sequence lengths (though the impact is small for
the longer sequence lengths and only substantial for
the short sequence lengths with $B=75\%$). 
Nevertheless, the impact on species tree estimation 
on these data tends to be
neutral, but there are also
conditions where binning was beneficial.

On the lower ILS 10-taxon datasets, 
statistical 
binning reduces gene tree estimation error, and
both weighted and unweighted binning reduce
species tree estimation error for $B=75\%$. 
However, species tree estimation error 
is unchanged when $B=50\%$.

The results on the higher ILS 10-taxon datasets
stand out from the other analyses:
statistical binning  slightly increases
gene tree estimation error 
when B=50\% but substantially increases gene
tree estimation error when B=75\%.
Furthermore, while species tree estimation error
is not increased for $B=50\%$, when
$B=75\%$, the error increases.

The difference in impact for statistical binning
in this case is interesting, and points out the
significance of how $B$ is set.
To understand this, note that when $B$ is very small, then
bin sizes will tend to be very small, since any pair of incompatible
branches with support above $B$ will be considered to be evidence of statistically significant 
discord; thus, small settings  for
$B$ produce results that are similar to unbinned
analyses. Conversely, very large settings for $B$ 
are more likely to bin genes together, since only
the strongest supported conflicting branches will prevent genes
from being binned. Therefore, 
if all the gene trees have low support
then statistical binning could tend to produce results
that are similar to concatenation.
Thus, the choice of the threshold matters.

To better understand the difference in impact of
statistical binning
on these simulated datasets, 
it is helpful to consider the ILS levels and gene tree bootstrap
support values for these
data. 
As shown in Table \ref{table:model-condition}, 
the average distance between the
true gene trees and the species trees
ranges for these datasets from as low as 18\%
(for the Mammalian 2X collection) to above
80\% (for the 10-taxon higher ILS collection and the 15-taxon
collection). 
Fig.~\ref{fig-fn-avian}  
shows how the effect of statistical binning used
with MP-EST
is impacted by ILS level
on the avian datasets: statistical binning
provided an improvement at all ILS levels, with
the largest improvement for the lowest ILS
level (2X branch lengths) and the smallest
improvement on the highest ILS level (0.5X branch
lengths). S3-S5 Figs.~evaluate
this issue on the mammalian datasets,
and shows large improvements provided by statistical
binning
under the lowest (2X branch lengths) ILS
level, smaller improvements under the middle (1X branch
length) ILS level, and then no improvement under
the highest (0.5X branch lengths) ILS level. 
Thus, statistical binning provided an
improvement except for
a small number of model conditions:
some of the 15-taxon conditions (which have 
discordance of 82\%), the higher ILS 10-taxon
conditions (which have discordance of 84\%),
and the highest ILS mammalian condition (which
have discordance of 54\%). 
S2 Table  
shows that the average bootstrap support
for the higher ILS 10-taxon datasets is
quite low -- only 37\%. Thus, 
statistical binning  seems to be beneficial when
both
ILS level and gene tree bootstrap support are not too high,  will
be neutral when bootstrap support values are high (so little or no 
binning occurs), but can be detrimental when 
ILS levels are extremely high but gene tree bootstrap support is low
enough that binning occurs. 
Thus, one consequence of
this study is the suggestion that when 
ILS levels are very high and the average gene tree
bootstrap support is low, then either statistical binning
should not be used, or it should be used in a very conservative
fashion -- with the parameter $B$ set very low.  


\section*{Conclusions}
Because species trees and gene trees can differ, the estimation of
species trees requires multiple loci.
One approach to estimating species trees from multiple conflicting
loci seeks to restrict the 
set of loci using principled arguments \cite{Salichos2013}, but
other approaches that explicitly model the discordance have
also been developed.
When gene tree discord is due to incomplete lineage sorting, then
summary methods, such as 
MP-EST or ASTRAL, can be used to estimate the species tree by combining gene trees. 
However, this study, as well as others
\cite{Leache2011,Patel2013,naive-binning,astral,GatesyMPE2014,MirarabStatBinning}, demonstrates
that gene tree estimation error impacts species tree estimation, so that
species trees estimated using summary methods  on  poorly estimated
gene trees can have low accuracy.
The (unweighted) statistical binning technique
proposed in \cite{MirarabStatBinning} improved the accuracy of estimated
gene trees, and was shown to improve
the accuracy of MP-EST when 
applied to MLBS gene trees.
However, as we proved here, 
using
unweighted statistical binning
within a phylogenomic pipeline
can be statistically inconsistent under the GTR+MSC model.
This is a significant issue. 

This study described a simple 
modification to statistical binning,  obtained by
replicating each supergene tree by the number of genes in its
bin (equivalently, replacing each gene tree from the input set by
its recalculated tree, which is the supergene tree for the bin). This 
modification, which we call ``weighted statistical binning" (WSB), 
is statistically consistent under GTR+MSC model
(i.e., as the number of genes and the number of sites for each
gene increases, the estimated species tree topology
converges to the true species tree topology),
and so addresses this drawback.
However, the current mathematical theory 
does not suggest any advantage will
be gained using
WSB within a phylogenomic pipeline, compared to 
an unbinned analysis (i.e., the use of
the summary method without binning) because when gene sequence length is unbounded,
unbinned analyses using summary methods are also statistically consistent.
Indeed, 
the current mathematical theory about standard coalescent-based summary methods
does not establish any guarantees
in the presence of gene tree estimation error (which is inevitable given
limited length), and
the same limitation applies to the theory established for WSB. 
Hence,
from a theoretical standpoint, there is no benefit obtained
in using WSB - at least not according to the current mathematical theory.

The reason to use WSB within a phylogenomic pipeline
is empirical -- how it
impacts the accuracy of the estimated species trees - and so
our study focused on whether WSB 
tends to increase or decrease the accuracy of summary methods, and
how the model conditions impact the relative 
performance of binned and unbinned analyses.

On the biologically-based simulated datasets, 
weighted and unweighted statistical binning generally
improved estimated gene tree distributions and
led to improvements for MP-EST and ASTRAL 
estimations of species tree topologies. 
The use of 
statistical binning with MP-EST
also improved estimated species tree branch lengths, 
increased bootstrap support for true positive edges, 
and reduced the number of
highly supported false positives, compared to unbinned MP-EST analyses. 
These improvements increased when
gene sequence alignments had low phylogenetic signal, 
the species tree had low ILS, or there were many genes.

The estimation of  species tree branch lengths is
biologically significant since these lengths are used to 
infer the amount of ILS in the data. Unbinned
MP-EST 
analyses 
tended to substantially underestimate branch
lengths (and thus over-estimate ILS), but both weighted and unweighted 
binning reduce this problem and produce branch lengths that are 
much closer to their true lengths. 
Since MP-EST tends to over-estimate ILS
in the presence of gene tree estimation error,
this means that predictions of ILS levels for
biological datasets may have been over-estimated.
Another consequence of this
observation is that the biologically-based model species trees used
here and in \cite{MirarabStatBinning,MirarabSysBio2014}  may have inflated
levels of ILS, since they used 
MP-EST to construct the model species tree.
If so, then performance under the lower ILS levels (species
tree branch lengths of 2X or larger)  might be closer to the
biological dataset conditions than the default 1X condition
and higher ILS conditions. 

The improvement in branch support is biologically
relevant, especially since unbinned MP-EST analyses sometimes produced
highly supported false positive branches in 
the presence of poorly estimated gene trees and low levels of ILS,
but binning reduced the incidence of these 
false positive branches with high support.


The results on small numbers of species, and in 
particular on the higher ILS 10-taxon datasets, show somewhat 
different trends. While results on the 15-taxon
datasets showed binning generally being helpful or neutral, 
statistical binning ranged from neutral
to detrimental  
on the higher ILS 10-taxon
datasets (however, the differences were not statistically significant).
Both the higher ILS 10-taxon and 15-taxon datasets had 
extremely high levels of ILS (the two highest
we examined -- average topological distance between
true gene trees and the true species trees 
of 84\% and 82\%, respectively). 
Given that statistical binning ranged 
from neutral to highly beneficial for all the other
model conditions, these data suggest that 
statistical binning may not be suitable to datasets with
extremely high ILS levels. 
Clearly, further
research is therefore needed to understand
the conditions under which binning will be beneficial
and where binning may reduce accuracy. 

This study also did not examine model conditions in which gene
tree estimation error is due to model misspecification, 
nor other biological causes for gene tree discord,
such as gene duplication and loss or horizontal gene transfer.
Furthermore, while we examined sequence datasets with 
varying numbers of sites for each locus (including some with 100bp),
even shorter sequences may be needed
to avoid loci that include any
recombination \cite{GatesyMPE2014}.

This study mainly examined
the impact of statistical binning on  MP-EST, and
examined its impact on ASTRAL only for a subset of
the data and only with respect to species tree topology
estimation (instead of the full set of criteria). Thus,
an important direction
for future study is to consider other
coalescent-based methods for estimating the species tree
from multiple loci. 
As a simple example, 
Mirarab \emph{et al.} \cite{MirarabSysBio2014} 
showed that the accuracy 
of MP-EST species trees depended on whether MLBS or best maximum likelihood
(BestML)
gene trees  were used, and that 
MP-EST trees based on 
BestML gene trees generally 
produced more accurate species tree topologies for datasets with large numbers of
genes (such as some of the model conditions studied in this paper).
The explanation offered for this is that BestML gene trees are
generally closer to the true gene tree than MLBS gene trees, and that this helps
coalescent-based species tree estimation. 
Hence, 
the evaluation of the impact of binning 
on MP-EST with BestML gene trees is  also needed.
It is also possible that
better results would be obtained
using Bayesian methods (such as MrBayes \cite{mrbayes}), rather than
MLBS, 
to generate the distribution of gene trees \cite{DeGiorgioDegnan2014},
since the posterior distribution produced by
Bayesian MCMC methods may 
be more closely centered around the true gene tree than
the MLBS sample.

This study suggests that
substantial improvement in species
tree estimation could be obtained if
we can develop more accurate methods for gene
tree estimation.
For example,
methods that co-estimate gene sequence alignments and trees, such as
BAli-Phy \cite{Redelings2005}, SAT\'e \cite{liu-science,sate-two}, and
PASTA \cite{pasta}, might provide improved 
gene tree estimation accuracy, 
compared to standard two-step procedures for
estimating trees (first align, and then compute the tree).

Indeed, another challenge is that {\em if} loci are restricted
to ultra-short sequences (10-50 sites), so as to decrease 
the probability of intra-locus recombination, then
approaches based on combining estimated gene trees
may not be able to provide highly accurate results, 
no matter what techniques are used to estimate gene trees.
Hence, 
it is also possible that  methods that construct species trees
directly from the sequence data,
rather than by combining gene trees, will have the best accuracy
(see, for example, \cite{ChifmanKubatko2014,Dasarathy2014,DasarathyNowakRoch2014}),
since they can avoid the analytical and empirical challenges caused
by gene tree estimation error.


However, as observed in this and other studies
\cite{Leache2011,GatesyMPE2014},
concatenation
often produces more accurate trees than
even the best coalescent-based methods when
the level of ILS is low enough.
Therefore, an important question 
is whether a given biological dataset has a sufficiently high 
level of ILS that a coalescent-based analysis is needed.
Conversely, 
coalescent-based methods that are
not only more accurate than concatenation
under conditions with high ILS, but also
comparably accurate even under low levels of ILS,
would be very helpful tools.

Finally, since statistical binning did reduce
accuracy for some of the data we examined
with small numbers of species and the very highest ILS levels, 
an important question that needs to be
addressed is 
whether these very high ILS
simulation conditions explored here and
elsewhere represent
realistic levels of ILS, or whether 
they represent extreme conditions that are unlikely to
be observed in nature. 
Accurate estimations of ILS levels in biological
data would enable the research community to
direct its efforts to developing methods that
would have the greatest utility in practice.

Overall, this study confirms the general
finding in \cite{MirarabStatBinning} 
that highly accurate coalescent-based species tree estimation
is possible, 
and that statistical binning used with good coalescent-based methods can 
provide improved accuracy relative to concatenation
under many conditions.

\section*{Materials and Methods}
\subsection*{Proofs}
Recall that under the GTR+MSC model,  gene trees evolve
within a species tree under the multi-species coalescent (MSC) model,
and then sequences evolve down each gene tree under the General
Time Reversible (GTR) model. The different gene trees
are equipped with their own GTR model parameters, and so
the tree topologies, $4 \times 4$ substitution matrices, and
gene tree
branch lengths
can differ between the different genes. 

The main results of this section are given in Theorem 2 
and Theorem 3,  
where we prove that using weighted statistical binning
in a phylogenomic pipeline is
statistically consistent under the GTR+MSC model,
but that replacing weighted statistical binning 
with unweighted statistical binning is {\em not} statistically
consistent under the GTR+MSC model, respectively.

The statistical binning
algorithm uses a heuristic to color
the vertices, which we now describe.
Since each gene is associated with a vertex, we will 
describe the heuristic in terms of what it does with genes.
The algorithm has two stages. 
In the first stage, we use a heuristic to find a large
clique in the incompatibility graph (i.e.,
a set of pairwise incompatible genes)  and we assign each gene in the 
clique to a different bin. 
Then, in the greedy stage, genes are processed in turn (according to an order described below), and
each gene is placed in the bin with the smallest number of genes with which it has
no strongly supported conflicts (where strongly supported
conflict between two genes
means that there are branches, one for each of the estimated gene trees,
that are incompatible, and both branches have support above $B$, where 
$B$ is the user provided bootstrap threshold support value).
If no such bin exists, a new bin is created and the gene is placed in this new bin. 
If there are two or more bins with the same smallest number of genes into which
the new gene can be placed, the tie is broken randomly.
Genes are processed based on a dynamic ordering, 
such that the next selected gene is always the one incompatible
with the largest number of existing bins (breaking ties arbitrarily).
Therefore, 

\vspace{.1in}
\noindent
 Lemma 1: 
Let $\mathcal{T} = \{t_1, t_2, \ldots, t_p\}$ be the multi-set of
estimated gene trees for $p$ genes $g_1, g_2, \ldots, g_p$, and assume that
all the branches in each $t_i$ 
have bootstrap support above $B$, the user-provided bootstrap support threshold.
Then,  when statistical binning is run, 
there will be one bin for each
of the different estimated gene tree 
topologies in $\mathcal{T}$,
and for every bin, every two genes in the bin will have the same
estimated gene tree topology.

\noindent
Proof: 
Our inductive hypothesis is that after placing $K$ genes into
bins, there will be one bin for each of the estimated gene tree topologies
for this set of $K$ genes, and that every two genes in any bin
will have the same estimated gene tree topology.
We will prove the lemma true by induction on $K \geq 1$.

For $K=1$, it is trivially true. Now suppose the inductive
hypothesis holds for $K-1$ genes, and consider what happens
when the $K^{th}$ gene, $g_K$, is placed. 
Recall that the algorithm operates in two stages: first it 
finds a clique in the graph and places the genes within that clique
into different bins, and then it enters the greedy phase.
We can consider the genes within the clique to be arbitrarily
ordered, and placed in bins using that order.
There are two cases to consider, depending on
whether $g_K$ is
part of the initial clique found by the algorithm.
If $g_K$ {\em is} part of the initial clique, then
$g_K$ is placed in a separate bin by itself, and has a different
topology from the $K-1$ genes that preceded it (because these are also
in the initial clique, and are placed in bins by themselves). 
If  $g_K$ is {\em not} part of the initial clique, 
then it is placed in a bin during the greedy stage of the algorithm.
By the inductive hypothesis,  
the algorithm has placed the first $K-1$
genes  into bins, there is a single 
bin for each of the different estimated gene tree topologies observed
among the first $K-1$ genes, and every two genes
in any bin have the same estimated gene tree topology.
When we process  $g_{K}$, there are 
two cases, depending on whether its estimated
gene tree  $t_{K}$ is a gene tree topology
that has been seen before. If $t_{K} = t_i$ for some $1 \leq i \leq K-1$, 
then there is a bin that contains all the
genes with that topology, and  $g_{K}$ can be added to that bin. 
Note that by the inductive
hypothesis, all other bins contain genes with different
estimated gene tree topologies than $t_{K}$. Furthermore,
by assumption,  all edges of all gene trees have
bootstrap support above $B$. Hence, 
we cannot add 
$g_{K}$ to any other bin. 
Therefore, if $t_{K}$ has
been seen before, there is only one bin we can add $g_{K}$ to,
and it is the bin for genes with the same tree topology as $t_K$.
The other case is where $t_{K}$ has not been seen before.
In this case, $t_{K}$ is different from every previously seen
gene tree, and so a new bin is created. As a result, the new set of 
bins satisfies the
inductive hypothesis,  so that there is one bin for every 
estimated gene tree topology,
and no two genes in any bin have different estimated gene tree topologies.
\QED

\vspace{.1in}
Therefore,


\vspace{.1in}
\noindent
Theorem 1: 
Let $T^{sp}$ be a species tree with branch lengths in
coalescent units, and 
$\mathcal{T} = \{t_1, t_2, \ldots, t_p\}$ be a set of $p$ rooted gene trees sampled from
the distribution defined by $T^{sp}$ under the multi-species coalescent model.
Let $\{\theta_1, \theta_2, \ldots, \theta_p\}$ be a set of numeric GTR model
parameters (gene tree branch lengths and $4 \times 4$ substitution
matrices) so that $T_i = (t_i,\theta_i)$ is a GTR model tree for
each $i=1,2,\ldots, p$.
Let $\mathcal{T}' = \{T_1, T_2, \ldots, T_p\}$.
For each $i, 1\leq i \leq p$, 
let sequence dataset $S_i$ 
evolve down the GTR model tree $T_i$.
Let $\epsilon < 1$ and bootstrap support threshold $B < 1$ 
be given.
Then, 
there is a sequence length $L$ (that depends on $\mathcal{T}'$ and $\epsilon$)
such that if at least $L$ sites evolve down each gene tree, then
with probability at least $1-\epsilon$, 
the following will be true:
\begin{itemize}
\item For each $i = 1, 2, \ldots, p$, the gene tree estimated using GTR maximum likelihood
on $S_i$ will have the same unrooted topology as $t_i$ (the true
gene tree for $S_i$), 
and will have bootstrap support greater
than $B$ for all its branches, 
\item 
For every bin produced by statistical binning based on GTR maximum
likelihood analyses of the gene sequence alignments, 
the estimated gene trees for genes in the bin will have the same 
topology, and
\item 
All genes with the same true gene tree topology will be in 
the same bin. 
\end{itemize}

\noindent
Proof: 
Since GTR maximum likelihood is statistically consistent for
sequences generated by GTR model trees, then 
for any $\epsilon' >0$,
there is a sequence length $L_i$ such that given 
sequence dataset $S_i$  with at least $L_i$ sites
 generated
on $T_i$, the GTR maximum likelihood tree topology
for $S_i$ is $t_i$ (i.e., the true gene tree) and has bootstrap
support greater than $B$, with probability at least $1-\epsilon'$. 
Letting $L = \max_i \{L_i\}$, it follows that
all estimated gene trees will be the true gene trees and
have bootstrap support greater than $B$ with probability at least 
$1-p\epsilon'$.
Therefore, when  $\epsilon' = \frac{\epsilon}{p}$ and the sequences
are all of length at least $L$, 
the result then follows by
Lemma 1. 
\QED

\vspace{.1in}
Fully partitioned GTR maximum likelihood:
In a  fully partitioned
GTR maximum likelihood analysis,
the input is a set of $p$ multiple
sequence alignments, $\{S_1, S_2, \ldots, S_p\}$.
These alignments are concatenated into a supermatrix,
$M$, 
in which  the locations where the different alignments 
begin and end are also noted. 
The maximum likelihood score
of a candidate tree $t$ (note that $t$ specifies only a topology and not also branch lengths)
for input $M$ is

\begin{equation} \label{eq:part}
score(t) = sup_{\Theta} \{\prod_i^p Pr(S_i|(t,\theta_i)): 
\Theta = \{\theta_1, \theta_2, \ldots, \theta_p\}\}
\end{equation}

Thus, $\Theta$ denotes
a set of GTR model parameters (branch lengths
and GTR substitution matrix) for each of the
parts within the 
concatenated alignment $M$.
We will refer to the tree topology 
that
achieves the optimal score under this fully partitioned
analysis as the solution to the fully
partitioned maximum likelihood
analysis of the concatenated matrix, understanding
that the numeric GTR parameters (branch
lengths and substitution matrices) are estimated independently
for each part of the alignment, and hence can differ
arbitrarily between parts.


\vspace{.1in}
\noindent
 Lemma 2: 
Let $S$ 
be
a set of 
taxa, and let 
$S_i$ be a set of DNA sequences for $S$,  with $i=1,2,\ldots,p$.
Suppose that tree topology $t$ is 
an optimal solution for
GTR maximum likelihood for each $S_i$ (allowing various GTR parameters for different $i=1,2,\ldots,p$). 
Then $t$ will be
an optimal solution to a fully partitioned GTR maximum likelihood analysis 
on a concatenation
of $S_1, S_2, \ldots, S_p$.

\noindent
Proof: 
Recall that in a fully partitioned GTR maximum likelihood analysis,
the maximum likelihood score of a given candidate tree $t$
with respect to a matrix $M$ 
under a fully partitioned ML analysis is given by Equation \eqref{eq:part}.
Suppose that the tree topology $t$ is an optimal solution to 
GTR maximum likelihood for each $S_i$  but not an optimal solution to
the fully concatenated GTR maximum likelihood
analysis. Then, 
for some tree $t' \neq t$, $ score(t') > score(t)$. Therefore,
for at least one $i$, 
$sup_{\theta}\{Pr(S_i|(t',\theta)\} > sup_{\theta}\{Pr(S_i|(t,\theta))\}$.
But then $t$ is not an optimal GTR maximum likelihood
tree topology for $S_i$, contradicting our assumption.
Therefore, if the maximum
likelihood analysis is performed
in a fully partitioned manner, then tree topology $t$ will be an optimal solution to
the GTR maximum likelihood analysis. 
{\QED}

\vspace{.1in}

\noindent
 Comments: 
The use of a fully partitioned 
analysis that enables different parameters for different partitions 
is critically important for the proof.
Consider, for example, the result that would be obtained
given a set of $p$ sequence alignments for $n$ species, of which $p-1$ of them
are constant (meaning all the sequences are identical across all the
species), but one sequence alignment, $S_p$,  is obtained by evolving
sequences down a GTR gene tree, $T$. 
In a fully partitioned GTR maximum likelihood analysis, the $p-1$ multiple
sequence alignments
that exhibit no changes do not impact the
solution to maximum likelihood, 
because they have the same score for every
possible tree topology.  Therefore,  the outcome of
a fully partitioned  GTR maximum likelihood analysis of the concatenated alignment
will simply have the GTR  maximum likelihood
tree topology for $S_p$ (recall that a fully partitioned analyses does not produce one
unique set of branch lengths or other model parameters).
However, in an {\em unpartitioned} GTR maximum likelihood analysis, 
the result can be quite different -- because the
$p-1$ alignments without changes on them will drive down the
estimated branch lengths, which are held in common across all
the sites.  
See \cite{RochSteel-inconsistent} for an analysis of
the theoretical properties of 
unpartitioned maximum likelihood in the 
context of the multi-species coalescent model.
We now consider the result of applying weighted statistical binning
within a phylogenomic pipeline.

\vspace{.1in}
\noindent
 Corollary 1: 
Let $\mathcal{G} = \{g_1, g_2, \ldots, g_p\}$ be a set
of $p$ genes, and $T_i=(t_i,\theta_i)$ be the true gene tree and
GTR parameters (including branch length) for $g_i$, 
$i=1,2,\ldots,p$.
Let $B<1$ be the user provided
bootstrap support value.
Assume that the gene sequence alignment $S_i$ evolves down
the GTR model tree
$T_i = (t_i,\theta_i)$,
 for $i=1,2,\ldots, p$. 
As the sequence lengths for all the
genes increase 
then with probability converging to $1$, 
for each bin produced during a statistical binning analysis, 
the estimated gene trees will be the true gene trees, 
all genes in any bin will have the same estimated and true gene tree, 
and the supergene trees produced for each bin
will converge in probability to  the
common
true gene tree for the genes in the bin.

\noindent
 Proof: 
By Theorem 1, 
as the sequence length increases,
then with probability converging to $1$, 
the genes in each bin will 
share a common true gene tree topology, 
their estimated gene trees will be topologically identical
to each other and to the true gene tree, and will
each have bootstrap support greater than $B$.
By Lemma 2, 
under these conditions, 
a fully partitioned GTR maximum
likelihood analysis of  the concatenated
alignment of the genes in a bin  will produce
the true gene tree topology for 
the genes in the bin.
\QED

\vspace{.1in}

We now address the statistical consistency of
phylogenomic pipelines that use
weighted and unweighted statistical binning.

\vspace{.1in}
\noindent
 Theorem 2: 
The phylogenomic pipeline that
uses GTR maximum likelihood to estimate gene trees,
uses weighted statistical binning
to compute supergene trees, 
and then combines the supergene trees using
a coalescent-based summary method,
is statistically consistent under the 
GTR+MSC model.

\noindent
Proof: 
We begin with the proof of statistical consistency
for weighted statistical binning.
By Corollary 1, 
as the sequence length for each
gene goes to infinity ($k \rightarrow \infty$) 
all genes put in any bin by statistical binning 
will have the same true gene tree with probability converging to 1,
and the supergene trees produced for each bin will converge
in probability to this common true gene tree. 
In weighted statistical binning, this common true gene tree topology is
replicated as many times as the number of genes in the bin, 
and hence the distribution produced using weighted statistical
binning is identical to the distribution of the unbinned true gene trees.
Therefore, as both $k$ and $p$ increase,
the gene tree distribution produced by weighted statistical
binning converges to the true gene tree distribution.
The statistical consistency of the pipeline follows from
the use of a coalescent-based summary method, since
as $p \rightarrow \infty$, the species tree produced by 
the summary method given true gene trees converges to the true 
species tree. 
{\QED}

\vspace{.1in}
We now consider the case where we use unweighted statistical binning
instead of weighted statistical binning.

\vspace{.1in}
\noindent
 Theorem 3: 
The phylogenomic pipeline that
uses GTR maximum likelihood to estimate gene trees,
uses unweighted statistical binning
to compute supergene trees,
and then combines the supergene trees using
a coalescent-based summary method,
is statistically inconsistent under the
GTR+MSC model.

\noindent
Proof: 
The proof for  Theorem 2 
shows  that as the sequence length $k$ increases,
the set of
bins produced by statistical binning converges in probability
to having one bin for each of the true gene trees, and the
supergene tree for each bin converges to the common true gene tree for the bin.
As $p \rightarrow \infty$, the set $\mathcal{T}$ converges in probability to the
set of all possible gene trees (since
 all gene trees have strictly positive probability under the multi-species
coalescent model).
Hence, the multi-set of supergene trees produced by unweighted statistical binning will
converge to the set that has each possible gene tree appearing exactly once.
This is a flat distribution, and it is not possible to reconstruct the
species tree from a flat distribution. 
Hence, the use of unweighted statistical binning in a phylogenomic
pipeline is not statistically consistent.
\QED

\subsection*{Evaluation}
We explored the performance of MP-EST and ASTRAL
with weighted and unweighted statistical binning,
and also without binning. We also examine  concatenation 
of the entire set of gene sequence alignments
using an unpartitioned
maximum likelihood analysis using RAxML. 
We explore performance 
on a collection of simulated and biological datasets originally
studied in \cite{MirarabStatBinning}.
We applied MP-EST and ASTRAL to
 a set of RAxML  gene trees computed on
bootstrap replicates of each gene sequence alignment. 
With bootstrap ML gene trees for each gene, 
summary methods were applied with the site-only multi-locus 
bootstrapping (MLBS) procedure~\cite{seo2008}, implemented as follows. For each gene or supergene, 200 replicates of bootstrapping are performed using RAxML. Next, 200 replicates ($R_1, R_2,\ldots, R_{200}$) of input datasets to the summary methods are created such that $R_i$ contains the $i^{th}$ bootstrap tree across all genes/supergenes. The summary methods are then run on these 200 input replicates, and 200 species trees are estimated. Finally, the greedy consensus 
tree of these 200 estimated species tree is computed, and support values are drawn on the branches of the greedy consensus tree by counting the occurrences of each bipartition in the 200 species trees.

\subsubsection*{Triplet gene tree distribution error}
MP-EST computes species trees using the estimated distribution on 
rooted triplet trees defined by its input of gene trees. We therefore 
evaluated the impact of binning on the estimated gene tree distribution,
measuring the divergence between the triplet distribution of estimated gene trees and the triplet distribution of true gene trees. We represent the gene tree distribution by the frequency of each of the three
possible alternative topologies for all the $n\choose 3$ triplets of taxa, where $n$ is the number of taxa. Therefore, we have $n \choose 3$ true triplet distributions.
Hence, for each triplet of taxa, we have estimated triplet distributions using
the unbinned analysis, as well as 
weighted and unweighted binning analyses.
 We computed the Jensen-Shannon divergence of each of these $n \choose 3$ triplet 
distributions and showed the empirical cumulative distribution of these divergences. 
The Jensen-Shanon divergence is a symmetrized  and smoothed version of Kullback-Leibler divergence~\cite{kldiv} between two distributions $P$ and $Q$, and can be calculated as follows~\cite{fuglede2004jensen}:


\begin{equation}
\centering
 JS(P,Q)= \frac{1}{2}KL(P,M) + \frac{1}{2}KL(Q,M)
\end{equation}
where $M = \frac{P+Q}{2}$, and KL is the Kullback-Leibler divergence.

\subsubsection*{Species tree estimation error and branch support}
We compared the estimated species trees to the model (i.e., true) species tree (for the simulated datasets) 
or to the scientific literature (for the biological datasets).  
We measure topological error using the missing branch rate (also known as 
the false negative (FN) rate), which is the proportion of 
branches in the true tree that are missing from the estimated tree. 
We also reported the error in species tree branch lengths estimated by MP-EST
using the  ratio of estimated branch length to true branch length for those 
branches of the true tree that appear in the estimated tree;
thus,  1 indicates correct estimation, values above $1$ indicate
lengths that are too long, and values below $1$
indicate branch lengths that are too short. Note that
species tree branch lengths reflect the expected amount of ILS, and so
under-estimation of species tree branch lengths means 
over-estimation of ILS, and 
over-estimation of branch lengths means under-estimation of ILS.
We also computed the branch support 
of the false positive (FP) and true positive (TP) edges, where
false positive edges are present in the estimated tree but not in the true tree,
and edges that are present in both the estimated and true tree are true positive edges. 


\subsection*{Simulated datasets}
We studied 
four collections of 
simulated datasets: two 
based on biological datasets  that
were generated in a prior study
\cite{MirarabStatBinning}, and two new collections
with smaller numbers of species.
We briefly describe the simulation protocol for the biological
datasets, and 
direct the reader to \cite{MirarabStatBinning} for full details.

\subsubsection*{Mammalian simulated datasets}
This dataset was generated by \cite{MirarabStatBinning}, and
studied there and also in \cite{MirarabSysBio2014}. Here we describe
the procedure followed by \cite{MirarabStatBinning} to generate these data.
First, a species tree was computed for
the  full 
biological dataset in \cite{Song2012}, using
MP-EST (this was done before removing 23 erroneous genes),
and the tree topology and branch
lengths were used as the model tree.  
Thus, the  mammalian simulation model tree has an ILS level
based on an MP-EST analysis of the biological mammalian dataset.
Gene trees were simulated within this 
species tree under the multi-species coalescent model, and then the
branch lengths on the gene trees were defined using the 
gene trees estimated on the biological dataset.

Variants of the basic model condition
were generated by varying the amount of ILS, the number of
genes, and the sequence
length for each gene; these modifications also
impact the amount of gene tree estimation error and the
average bootstrap support in the estimated gene trees,
and so can be modified to produce datasets that resemble the
biological data.


The amount of ILS was varied by adjusting the branch length (shorter
branches increase ILS). A model condition with reduced ILS was created by uniformly
doubling (2X) the branch lengths, and a model condition with higher
ILS was generated by uniformly dividing the branch lengths by two (0.5X).
The amount of ILS obtained without adjusting the branch lengths
is referred to as ``default ILS'', and was estimated by MP-EST on the biological data.

The average bootstrap support (BS) in the biological data
was 71\%, and so \cite{MirarabStatBinning} generated sequence lengths
that produced estimated gene trees with bootstrap support
bracketing that value -- 500bp alignments produced estimated
gene trees with 63\% average BS and 1000bp alignments produced
estimated gene trees with 79\% BS. 
We also generated model conditions with very short sequence lengths (250bp), which have 43\% average BS.


Here, we varied the number of genes from 50 to 800
to explore both smaller and larger numbers of genes
than the biological dataset (which had
roughly 400 genes). In total, 
we generated 17 different model conditions specified by the ILS level,
the number of genes, and the sequence length. For each of these model conditions, \cite{MirarabStatBinning}
created 20 replicates.

\subsubsection*{Avian simulated datasets}

Mirarab et al.~\cite{MirarabStatBinning} used the species tree estimated by MP-EST on 
a subset of the avian dataset
with 48 species and 14,446 loci studied by  \cite{JarvisScience2014}, and simulated gene trees by varying different parameters (similar to the mammalian simulated datasets). Three types of genomic markers were studied in \cite{JarvisScience2014}:
exons, UCEs, and introns.
The average bootstrap support (BS) of the gene trees based on 
exons,  UCEs, and introns,  was 24\%, 39\% and 48\%, respectively; 
the longest introns had the highest average BS (59\%). 
Mirarab et al.~varied sequence lengths (250bp, 500bp, 1000bp, and 1500bp)
to produce four model conditions with
patterns of average bootstrap support that resemble these four marker types. 
Mirarab et al.~varied the number of genes from 200 to 2000, but 
here, we augmented the dataset to also look at fewer genes (50 and 100). 
Mirarab et al.~varied the amount of ILS, using the same technique as was
used in generating the mammalian simulated datasets.

\subsubsection*{15-taxon simulated datasets}
We simulated a collection of 15-taxon datasets.
The model species tree is a caterpillar-like
ultrametric tree (i.e., the substitution
process obeys a strict molecular clock)
with 15 taxa; hence, it has two leaves $x$ and $y$ that
are siblings in the tree.
The lengths of
all internal branches and the two branches incident with leaves $x$ and
$y$ 
are all set to 0.005 substitutions per site; note
that the assumption of ultrametricity defines 
the remaining branch lengths.
The population size parameter ($\theta=4N\mu$) is set to 0.05 for all branches,
and this results in
12 short internal branches (0.1 in coalescence units)
in succession.
Ultrametric gene trees were simulated down this 
tree using McCoal\cite{MCCoal-software} 
and commands given in S13 Fig. 
Sequence data were simulated down each gene tree using bppseqgen \cite{Dutheil2008}
according to GTR+$\Gamma$ parameters given in 
S13 Fig. 
We built four model conditions (with ten
replicates each) by trimming gene data to 100 or 1000 sites 
and by exploring 100 or 1000 genes.

\subsubsection*{10-taxon simulated datasets}
 We used simPhy \cite{simphy} to simulate species trees
 using the Yule process with two different maximum tree length settings:
 200K generations, resulting in short trees and high levels of ILS,
 and
 1.8M generations, resulting in relatively longer trees and lower levels of ILS. 
We generated 20 species trees per model condition, and used simPhy
to simulate 200 gene trees for each species trees using the multi-species coalescent process
(simPhy parameters and commands are given in S14 Fig.). 
The gene trees (with branch lengths in substitution units)
deviate from the strict molecular clock, and the
rates of evolution vary across genes.
We used Indelible to simulate GTR+$\Gamma$ sequence
evolution down these gene trees with 100 sites,
with parameters given in S14 Fig. 

\subsection*{Biological datasets}

We studied two biological datasets also studied in \cite{MirarabStatBinning}:
the avian dataset \cite{JarvisScience2014} containing 14,446 loci across 48 species, and a
reduced version of the
mammalian dataset studied by Song et al.~\cite{Song2012} with 447 loci across 37 species,
from which \cite{MirarabStatBinning}  deleted 23 erroneous genes and
re-estimated gene trees using RAxML
(see \cite{MirarabStatBinning,MirarabSysBio2014} for discussion of these loci). 

\subsection*{Methods and commands}

\paragraph{Gene tree estimation: }
RAxML version 7.3.5 \cite{Stamatakis2006} was used to estimate gene
trees under the GTRGAMMA model, using the following command:

\begin{itemize}
\item[]
  raxmlHPC-SSE3 -m GTRGAMMA -s [input\_alignment] -n
[output\_name] -N 20 \\ -p [random\_seed\_number] 
\end{itemize}

The following command was used for bootstrapping:

\begin{itemize}
\item[]
 raxmlHPC-SSE3 -m GTRGAMMA -s [input\_alignment] -n
[output\_name] -N 200 \\ -p [random\_seed\_number] -b
[random\_seed\_number] 
\end{itemize}

\paragraph{Supergene tree estimation: }

For the biological 
datasets and the 10- and 15-taxon simulated datasets,
we used a fully partitioned maximum likelihood
analysis. 
All other analyses were based on unpartitioned maximum
likelihood analysis, using
the command given above for 
gene tree estimation.  For
the fully partitioned analysis,  we used the following command:

\begin{itemize}
\item[]
 raxmlHPC-SSE3 -m GTRGAMMA -s [input\_alignment] -m GTRGAMMA -n
[output\_name] -N 20 \\ -M -q [partition\_file] -p [random\_seed\_number] 
\end{itemize}

\paragraph{Concatenation: }
\noindent We concatenate the alignments of all genes into one 
supermatrix, and then estimate a tree 
from the supermatrix using unpartitioned maximum likelihood. 
We computed a parsimony
starting tree using RAxML version 7.3.5, and then ran RAxML-light
version 1.0.6.  The following commands were used:

\begin{itemize}
\item[]
 raxmlHPC-SSE3 -y -s supermatrix.phylip -m GTRGAMMA
-n [output\_name] \\-p [random\_seed\_number] 

\item[]
 raxmlLight-PTHREADS -T 4 -s supermatrix.phylip -m
GTRGAMMA -n name \\-t [parsimony\_tree] 
\end{itemize}

\paragraph{MP-EST: }We used
version 1.3 of MP-EST. We ran MP-EST 10 times with different random
seed numbers, and selected the species tree with the best likelihood
score using a custom shell script.
MP-EST was run using site-only multi-locus bootstrapping, using 200
MLBS replicates, and returning the greedy consensus of the 200 MP-EST
MLBS species trees as the output. The branch support
on the edges of the tree represent the frequency of
the bipartition in the sample of 200 species trees.

\paragraph{ASTRAL: }We used ASTRAL version 4.7.6. in its default mode using the following command:
\begin{itemize}
\item[]
 astral.4.7.6.jar -i [input\_gene\_trees] -o [output\_file] 
\end{itemize}

\paragraph{Greedy consensus:}
The greedy consensus (also called the ``extended
majority consensus") of a set of trees, all on the same set of
leaves, is obtained by ordering the bipartitions that appear
in one or more trees in the order of their frequency (most frequent first).
Then, a tree is built from this set, beginning with the first bipartition, and then
modifying the tree to include the next bipartition in the list, if the addition of the bipartition
is possible. 
We used Dendropy version 3.12.0~\cite{dendropy} to compute greedy
consensus trees when running MP-EST or ASTRAL with MLBS gene trees.


\section*{Data Availability}
Most of the datasets used in this study are
available through the prior publications. The new
datasets generated for this study are 
available on figshare, with 
DOI: http://dx.doi.org/10.6084/m9.figshare.1411146.
(Retrieved May 13, 2015.)
The weighted statistical binning software is
available on github at \\
https://github.com/smirarab/binning
(Retrieved May 14, 2015.) 

\section*{Acknowledgements}
The authors acknowledge the support from
the Texas Advanced Computing Center (TACC), which
provided a generous allocation of
computing time so that these analyses could be performed.
 
\clearpage

\clearpage
\clearpage
\section*{Figures}
%

\clearpage

\beginsupplement     
\section*{Supporting Information}



\begin{table}[htbp]
\centering
\begin{tabular}{l|c|ccc}
\hline
Dataset & Model condition & \multicolumn{3}{c}{Gene tree error (\%)} \\ 
 & & Unbinned & Binned-50 & Binned-75 \\ \hline
      & 250bp & 79 & 57 & n.a. \\ 
Avian & 500bp & 69 & 57 & n.a.\\
      & 1000bp & 55 & 51 & n.a.\\
      & 1500bp & 46 & 45 & n.a. \\ \hline
          & 250bp & 60 & n.a. & 47 \\
Mammalian & 500bp & 43 & n.a. & 35 \\
          & 1000bp & 27 & n.a. & 26 \\ \hline
15-taxon & 100bp & 77 & 80 & 86 \\
         & 1000bp & 36 & 36 & 40 \\ \hline
10-taxon & Lower ILS & 64 & 58 & 51 \\
         & Higher ILS & 69 & 73 & 80 \\ \hline

\end{tabular}
\caption{\textbf{Gene tree estimation error, with and without binning for simulated datasets}. We show the average gene tree estimation error for the simulated datasets analyzed in this paper. Results are shown for fixed number of genes (1000 for avian
and 200 for mammalian, 100 for 15-taxon and 100 for 10-taxon). We fixed the level of ILS to 1X for avian, mammalian and 15-taxon datasets; and varied the level of ILS for 10-taxon datasets with 100bp sequence length. Gene tree error is mean topological distance, measured using the missing branch rate between the true gene tree and all 200 bootstrap replicates of each estimated gene tree. For the supergene trees, each bootstrap replicate of each supergene tree is compared separately against each true gene tree for the genes put in that bin. ``n.a.'' stands for ``not available''.}

\label{table-gt-error}
\end{table}


\begin{table}[htbp]
\centering
\begin{tabular}{l|l|cc}
\hline
Dataset & Model condition & Average bootstrap support (\%)\\ \hline
      & 250bp & 27 \\
Avian & 500bp &  31 \\
      & 1000bp &  51\\
      & 1500bp & 60  \\ \hline
          & 250bp & 43 \\
Mammalian & 500bp & 63 \\
          & 1000bp & 79  \\ \hline
15-taxon & 100bp & 35  \\
15-taxon & 1000bp &  70 \\ \hline
10-taxon & Lower ILS, 100bp & 45  \\
         & Higher ILS, 100bp & 37  \\ \hline

\end{tabular}
\caption{\textbf{Average bootstrap support}. We show the average bootstrap support values of the estimated gene trees for the simulated datasets. Results are shown for fixed number of genes (1000 for avian
and 200 for mammalian, 100 for 15-taxon and 100 for 10-taxon datasets). We fixed the level of ILS to 1X for avian, mammalian and 15-taxon datasets; and varied the level of ILS for 10-taxon datasets with 100bp sequence length.}
\label{table-gt-bootsupport}
\end{table}


\clearpage


\renewcommand{\thefigure}{S\arabic{figure}}
\setcounter{figure}{0}

\begin{figure}[!ht]
\begin{center}
\includegraphics[width=0.55\textwidth]{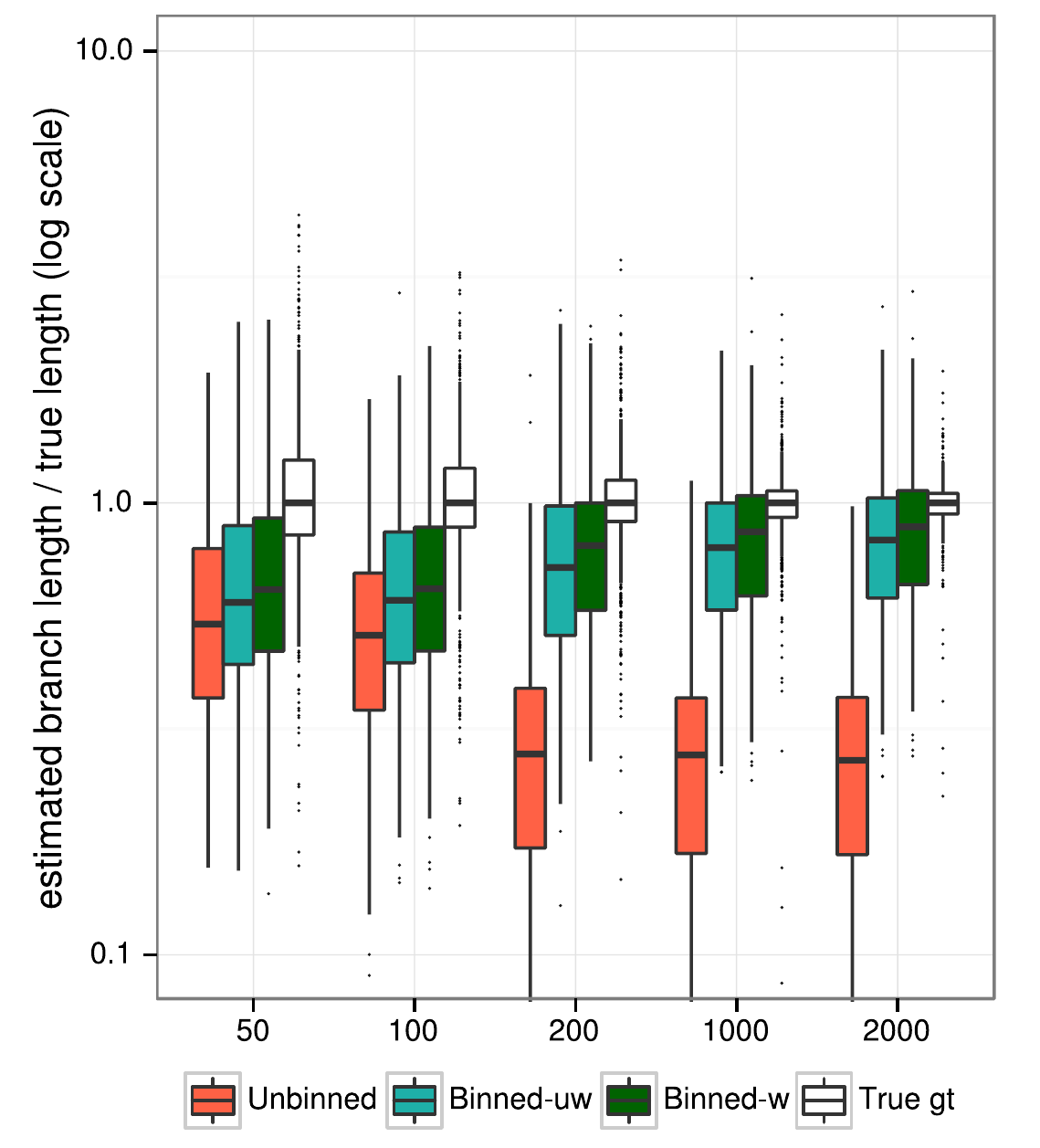}
\end{center}
\caption{{\bf Effect of binning on the branch lengths (in coalescent units) estimated by MP-EST using MLBS on the avian simulated datasets with varying numbers of gene trees}. We show the species tree branch length error (the ratio of estimated branch length to true branch length for branches of the true tree that appear in the estimated tree; 1 indicates correct estimation). We varied the number of genes from 50 to 2000, and fixed the sequence length to 500bp with default amount of ILS (1X level). We used 50\% bootstrap support threshold for binning. Supergene trees were estimated using unpartitioned analyses.
}
\label{fig:avian-gt-bl}
\label{fig6}
\end{figure}
%

\begin{figure}[h]
\begin{center}
\includegraphics[width=0.98\textwidth]{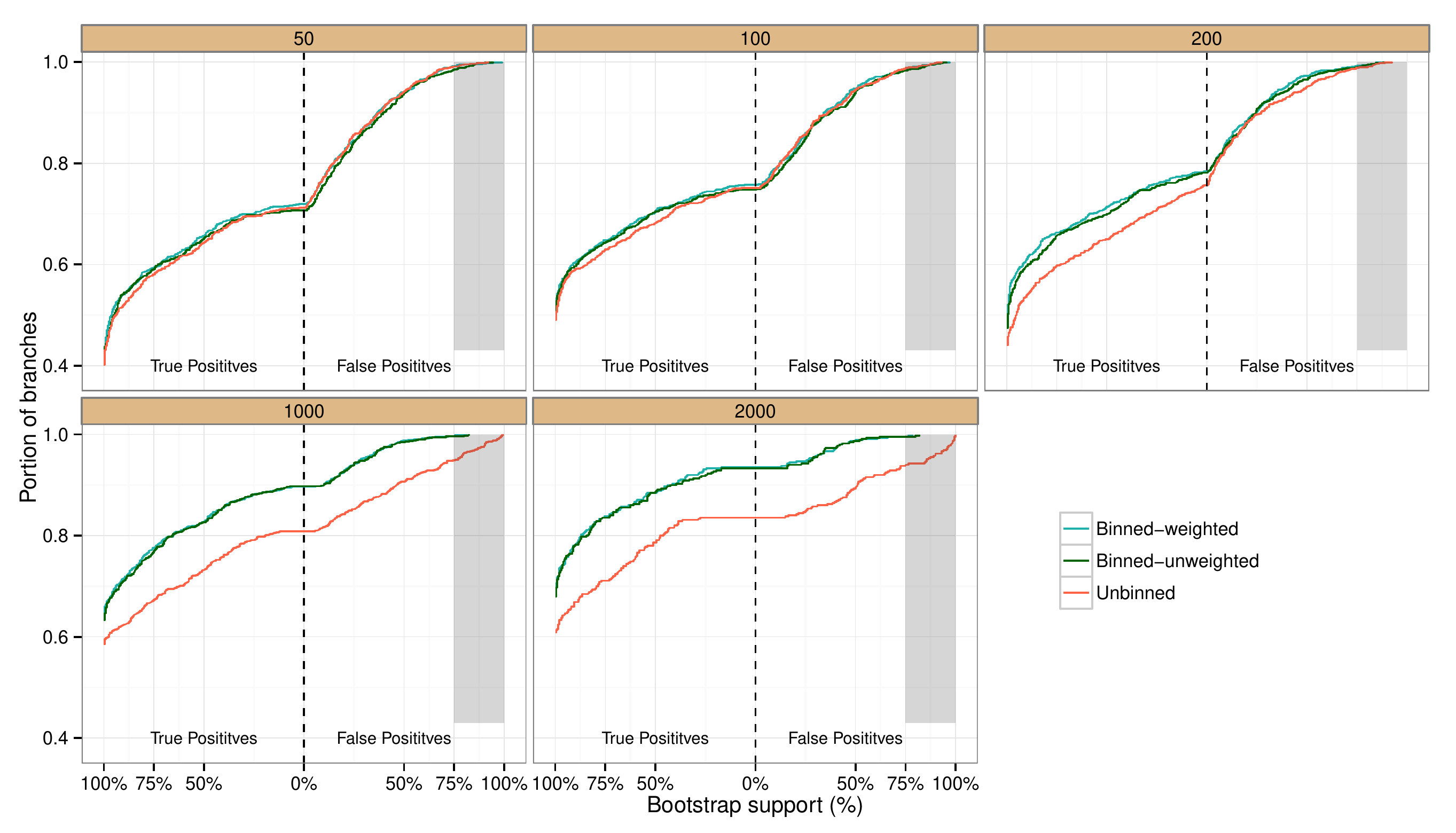}
\end{center}
\caption{{\bf Cumulative distribution of the bootstrap support values (obtained using MLBS) of true positive (TP) and false positive (FP) edges estimated by MP-EST on avian datasets}. We
varied the numbers of genes, and fixed the sequence length to 500bp (UCE-like)
with default amount of ILS (1X level). We used 50\% bootstrap support threshold for binning. Supergene trees were estimated using unpartitioned analyses. To produce the graph, we order the branches in the estimated species tree by their quality,
so that the true positives with high support come first, followed by lower support true positives,
then by false positives with low support, and finally by false positives with high support.
The false positive branches with support above 75\% are the most troublesome, and
that fraction are indicated in the grey area. When the curve for a method lies above the
curve for another method, then the first method has better bootstrap support.} \label{fig:tp-fp-avian-gt-mpest-ecdf}
\label{fig16}
\end{figure}


\begin{figure}[h]
\begin{center}
\includegraphics[width=0.98\textwidth]{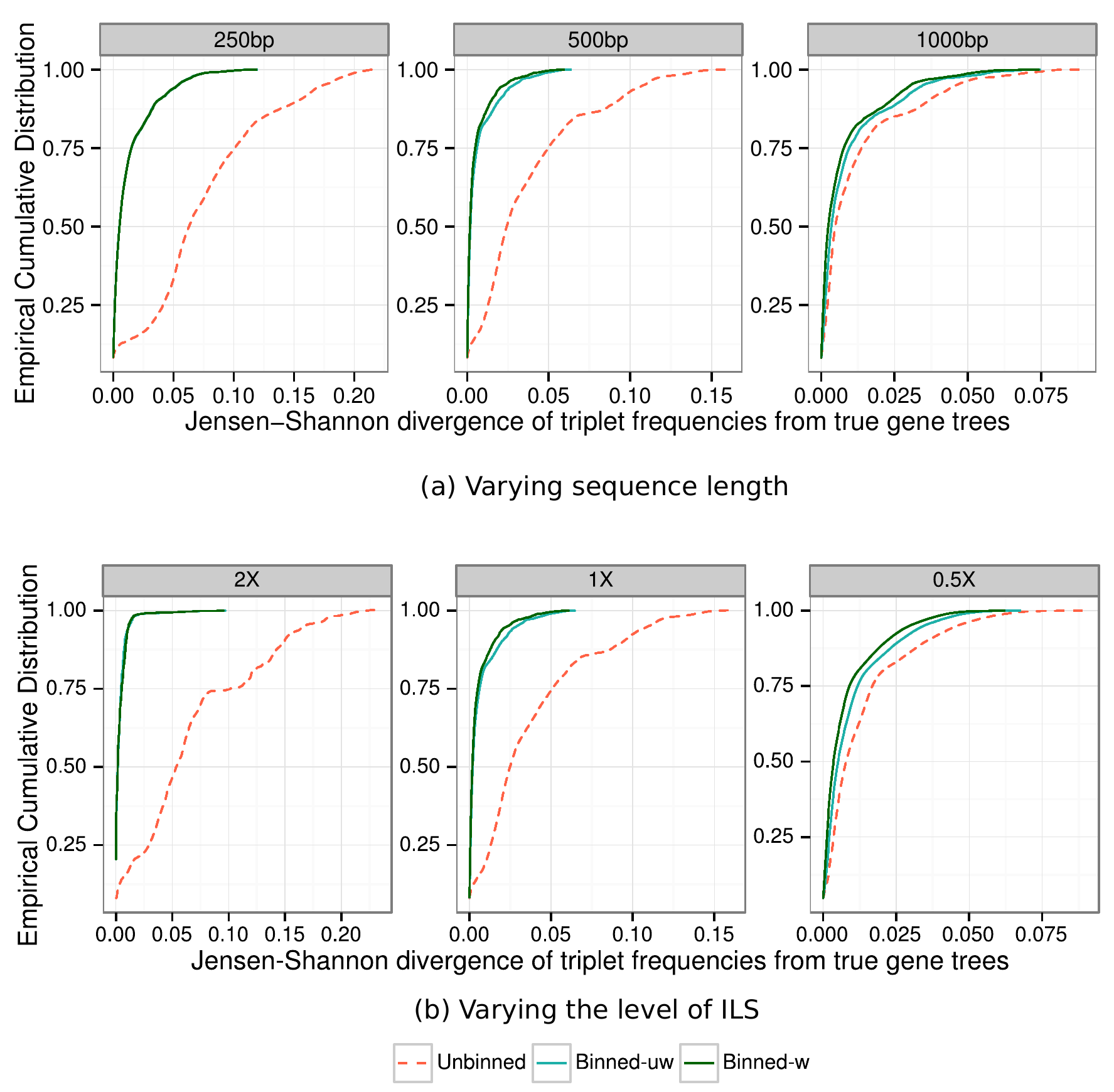}
\end{center}
\caption{{\bf Divergence of estimated gene trees triplet distributions from true gene tree
distributions for simulated mammalian datasets}. 
(a) Varying  gene sequence alignments lengths with 200 number of genes and default levels of ILS (1X); (b) varying ILS levels with fixed 200 genes and sequence length fixed to 500bp (63\% BS). We used 75\% bootstrap support threshold for binning. Supergene trees were estimated using unpartitioned analyses. True triplet frequencies are estimated based on true gene trees for each of the $n \choose 3$ possible triplets, where $n$ is the number of species. Similarly, triplet frequencies are calculated from estimated gene/supergene trees. For each of these $n \choose 3$ triplets, we calculate the Jensen-Shannon divergence of the estimated triplet distribution from the true gene tree triplet distribution. We show the empirical cumulative distribution of these divergences. The empirical cumulative distribution shows that for a given divergence level, what percentage of the triplets are diverged from true triplet distribution at or below that level. Results are shown for 10 replicates.
} \label{fig:mammal-shanon-bp}
\label{fig16}
\end{figure}

%

\begin{figure}[!ht]
\begin{center}
\includegraphics[width=0.45\textwidth]{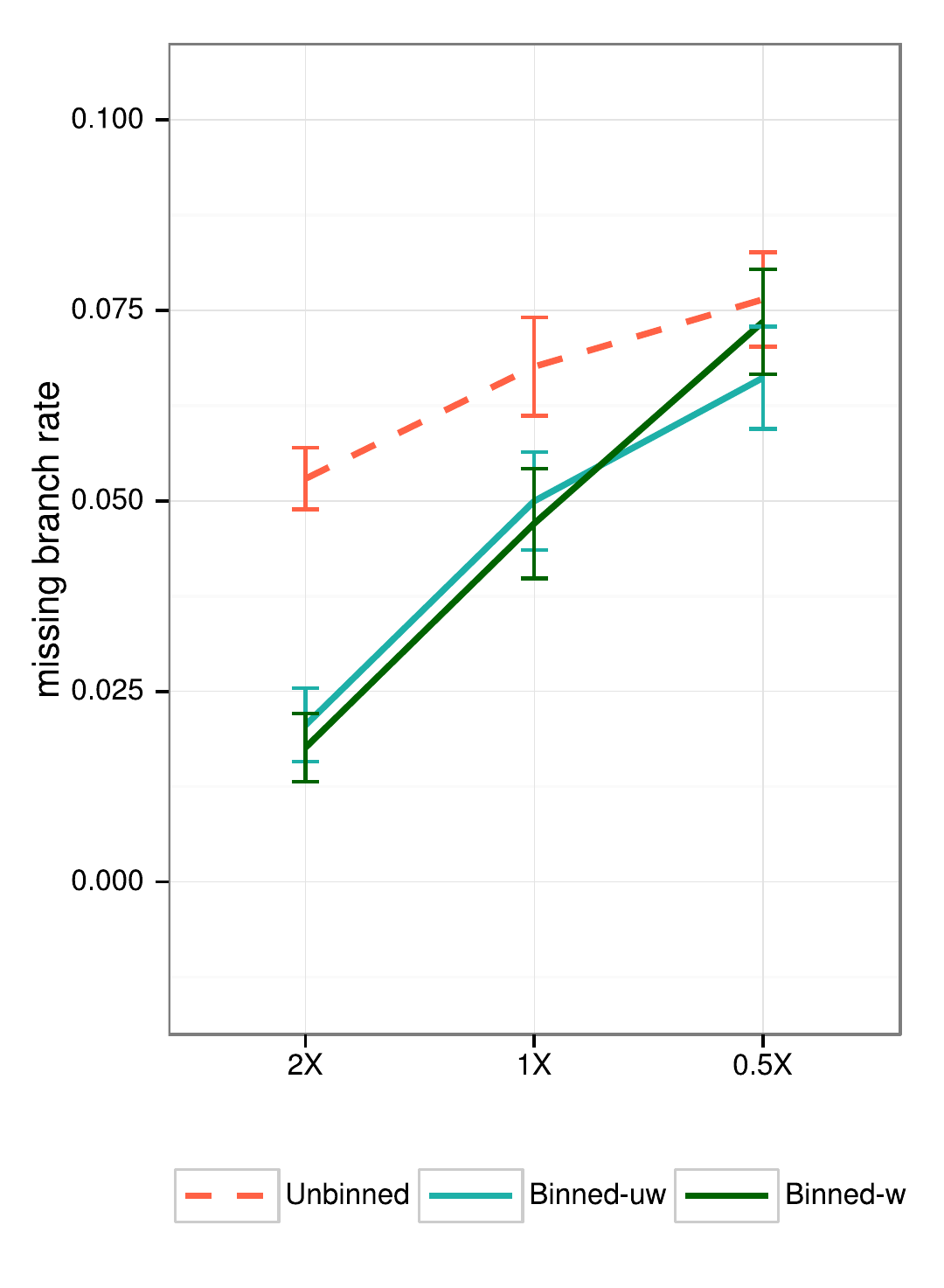}
\end{center}
\caption{{\bf Species tree estimation error for MP-EST with MLBS on
mammalian simulated datasets with varying amounts of ILS}. We show
average FN rate over 20 replicates. We varied the amount of ILS,
and fixed the number of genes to 200 and sequence length to 500bp
(63\% BS). We used 75\% bootstrap support threshold for binning. Supergene trees were estimated using unpartitioned analyses.} \label{fig:mammal-ILS-mpest}
\label{fig14}
\end{figure}


\begin{figure}[!ht]
\begin{center}
\includegraphics[scale=0.70]{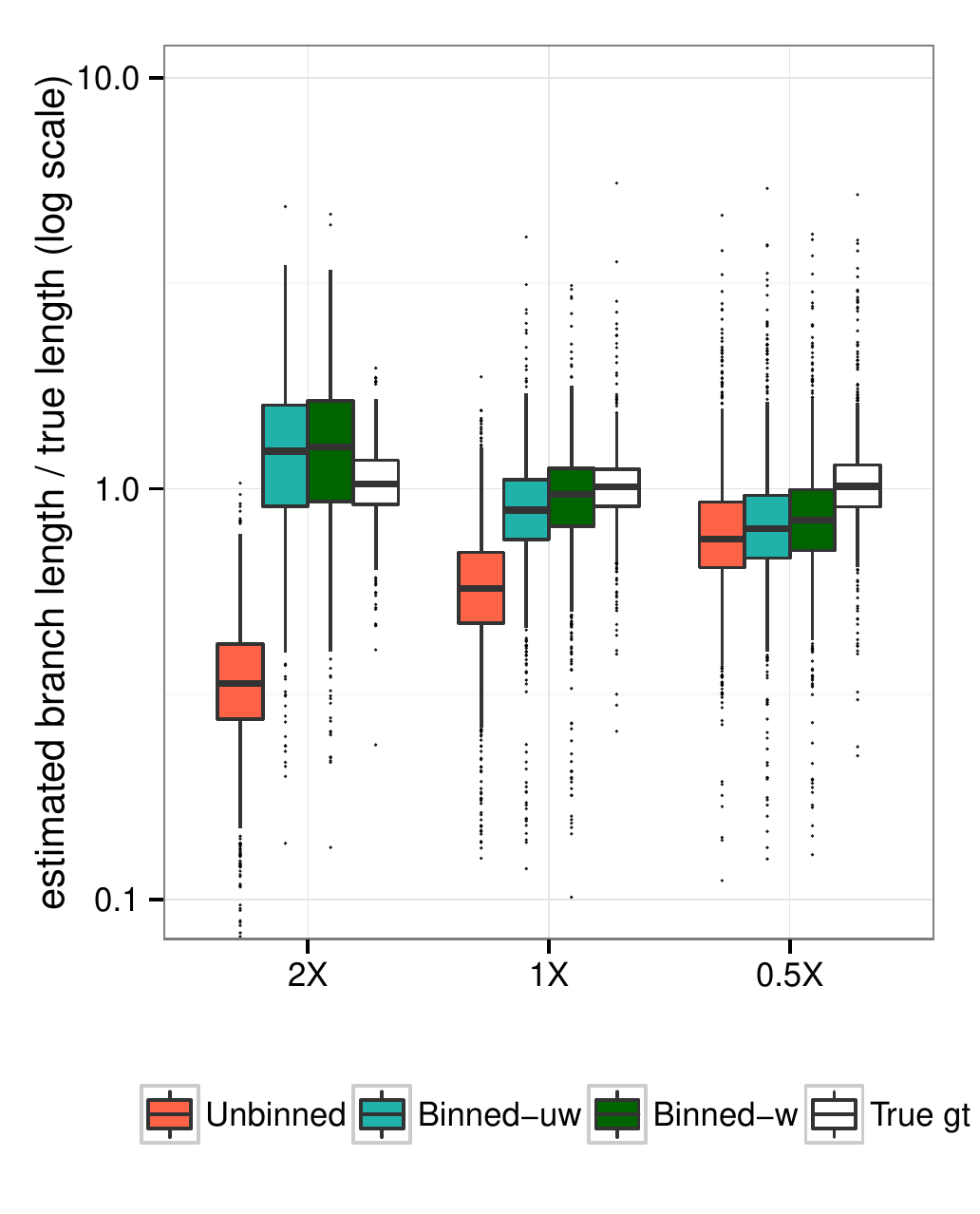}
\end{center}
\caption{{\bf Effect of binning on the branch lengths (in coalescent unit) estimated by MP-EST using MLBS on the mammalian simulated datasets with varying amounts of ILS}. We show the species tree branch length error (the ratio of estimated branch length to true branch length for branches of the true tree that appear in the estimated tree; 1 indicates correct estimation). We varied the amount of ILS, and fixed the number of genes to 200 and sequence length to 500bp (63\% BS). We used 75\% bootstrap support threshold for binning. Supergene trees were estimated using unpartitioned analyses.
}
\label{fig:mammal-bl-ILS}
\label{fig9}
\end{figure}


\begin{figure}[h]
\begin{center}
\includegraphics[width=0.98\textwidth]{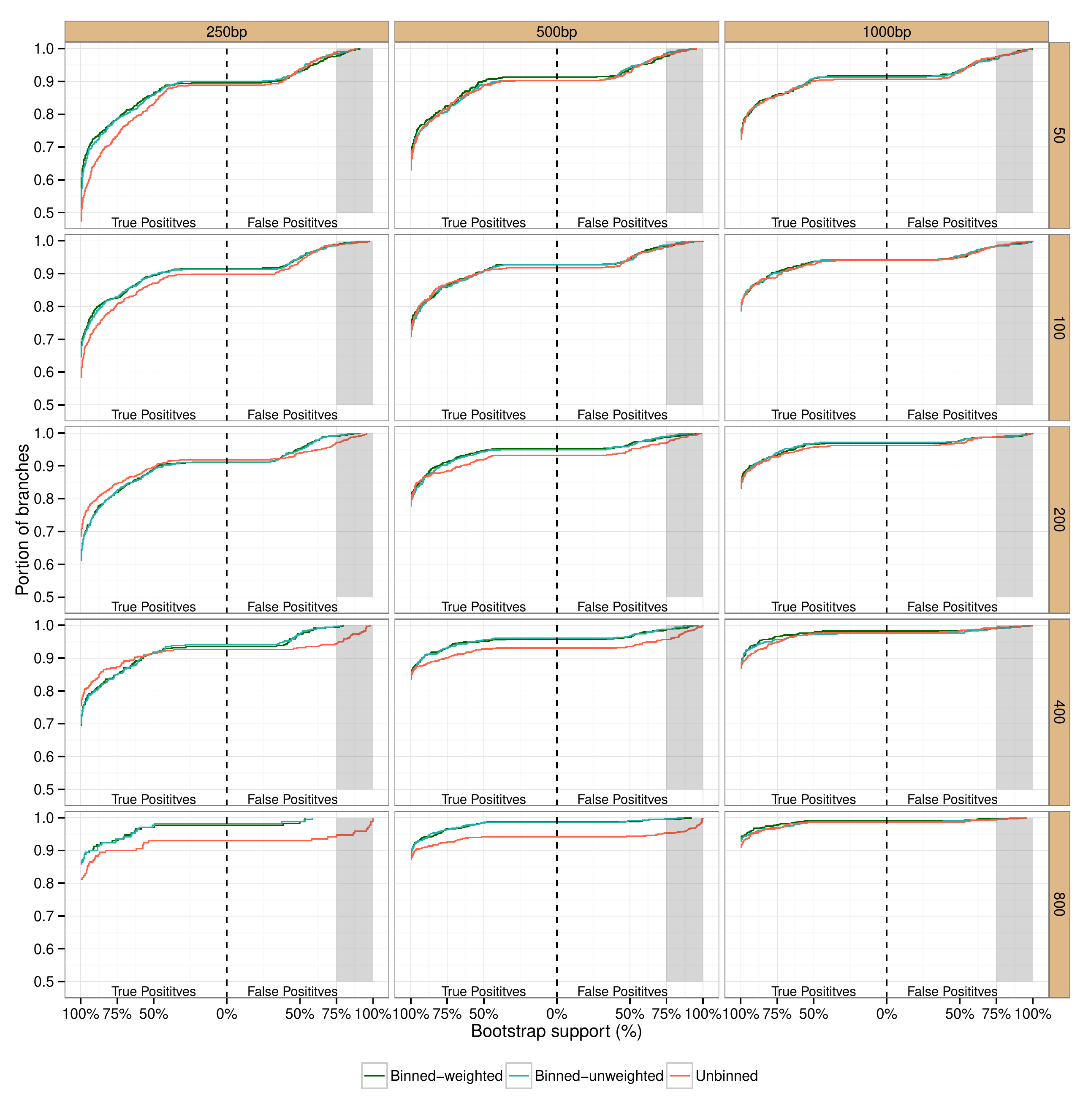}
\end{center}
\caption{{\bf Cumulative distribution of the bootstrap support values (obtained using MLBS) of true positive (TP) and false positive (FP) edges estimated by MP-EST on mammalian datasets}. We
varied the numbers of genes, and gene sequence alignments length with default amount of ILS. We used 75\% bootstrap support threshold for binning. Supergene trees were estimated using unpartitioned analyses.
To produce the graph, we order the branches in the estimated species tree by their quality,
so that the true positives with high support come first, followed by lower support true positives,
then by false positives with low support, and finally by false positives with high support.
When the curve for a method lies above the
curve for another method, then the first method has better bootstrap support.
} \label{fig:tp-fp-mammal-bpgt-mpest-ecdf}
\label{fig18}
\end{figure}


\begin{figure}[h]
\begin{center}
\includegraphics[width=0.98\textwidth]{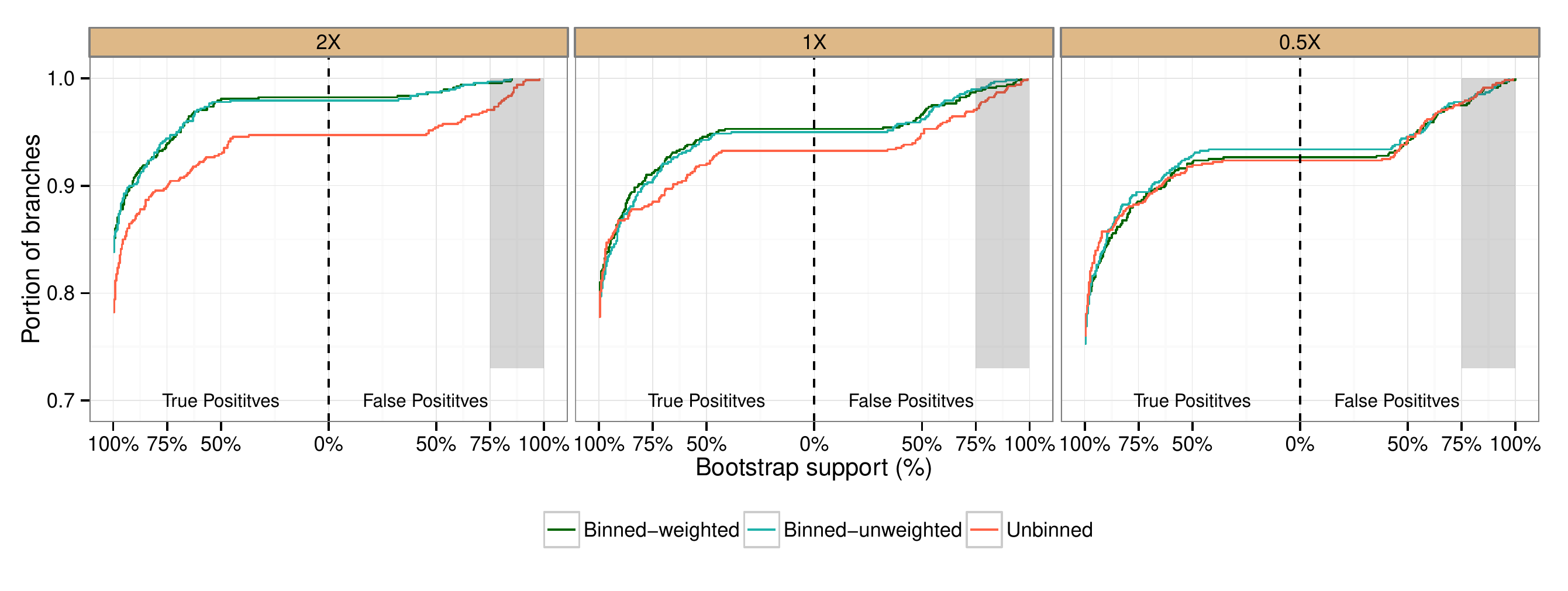}
\end{center}
\caption{{\bf Cumulative distribution of the bootstrap support values (obtained using MLBS) of true positive (TP) and false positive (FP) edges estimated by MP-EST on mammalian datasets with varying amounts of ILS}. We
varied the amount of ILS, and fixed the number of genes to 200 and sequence length to 500bp. We used 75\% bootstrap support threshold for binning. Supergene trees were estimated using unpartitioned analyses. To produce the graph, we order the branches in the estimated species tree by their quality,
so that the true positives with high support come first, followed by lower support true positives,
then by false positives with low support, and finally by false positives with high support.
When the curve for a method lies above the
curve for another method, then the first method has better bootstrap support.} \label{fig:tp-fp-mammal-ILS-mpest-ecdf}
\label{fig19}
\end{figure}


\begin{figure}[!ht]
\begin{center}
\includegraphics[width=0.75\textwidth]{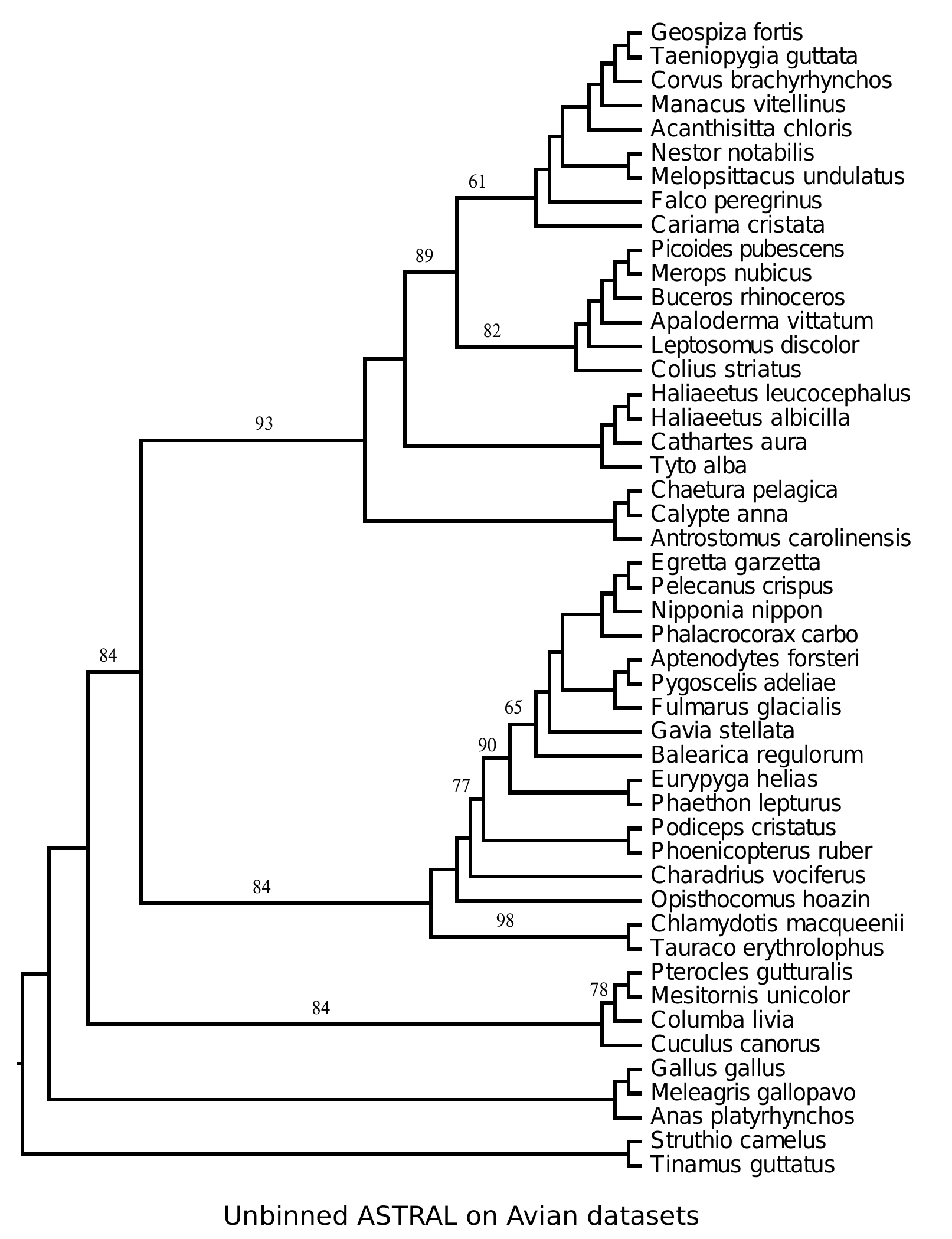}
\end{center}
\caption{Species trees estimated by unbinned ASTRAL using MLBS on avian biological datasets. Branches without
designation have 100\% support. We used 50\% bootstrap support threshold for binning. Supergene trees were estimated using fully partitioned analyses.} 
\label{fig:avian-bio-astral}
\end{figure}

\begin{figure}[!ht]
\begin{center}
\includegraphics[width=0.98\textwidth]{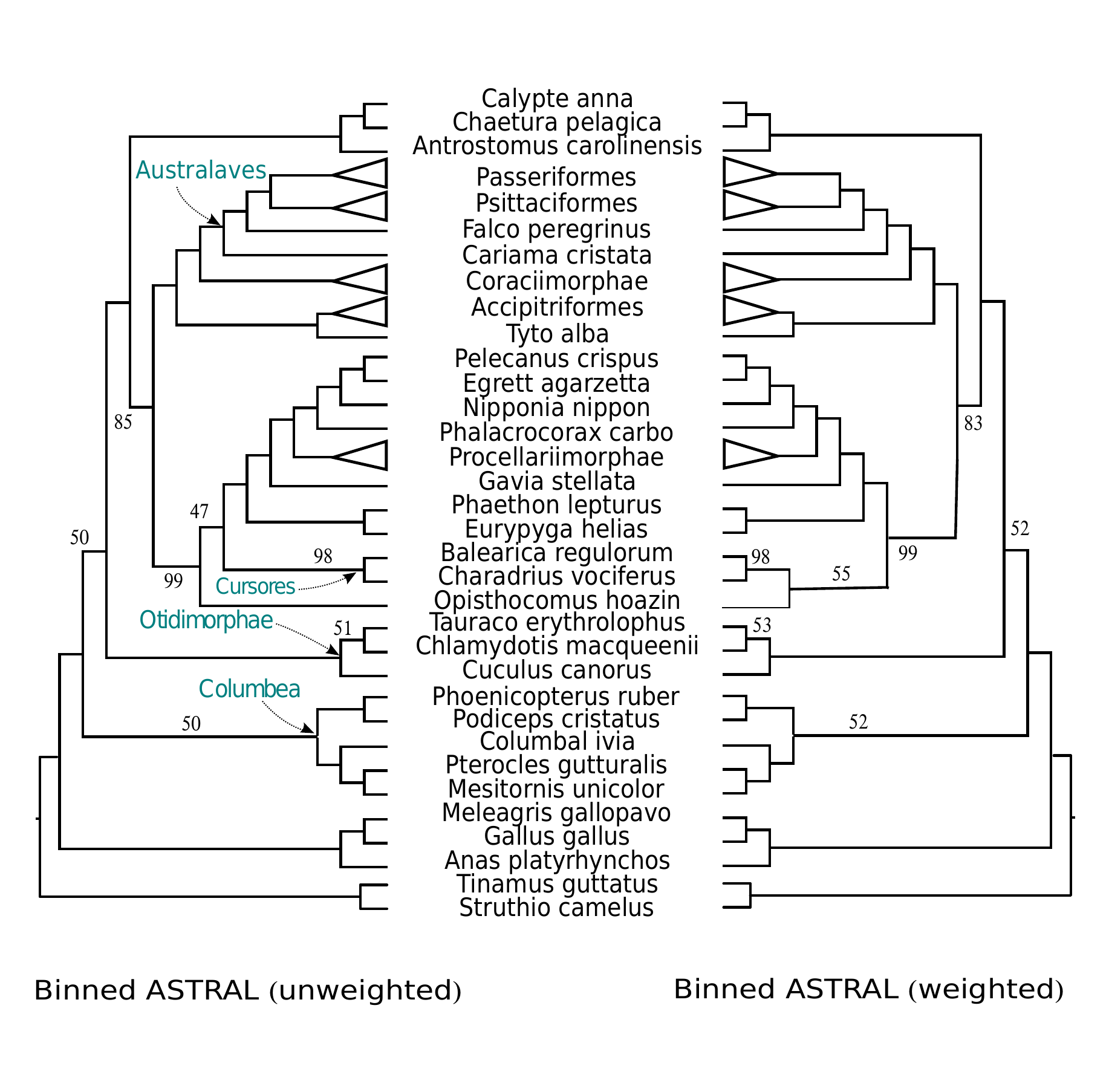}
\end{center}
\caption{Species trees estimated by binned (with and without weighting) ASTRAL using MLBS on avian biological datasets. (a) Unweighted binned ASTRAL, and (b) weighted binned ASTRAL. Branches without
designation have 100\% support. We used 50\% bootstrap support threshold for binning. Supergene trees were estimated using fully partitioned analyses. Binned ASTRAL with weighting and 
binned ASTRAL without weighting differ only in the placement of \textit{Opisthocomus hoazin}. However, the 
branches supporting different placements of \textit{Opisthocomus hoazin} have low support values (47\% for 
unweighted binning and 55\% for weighted binning).
} \label{fig:avian-bio-astral}
\end{figure}


\begin{figure}[!ht]
\begin{center}
\includegraphics[width=0.75\textwidth]{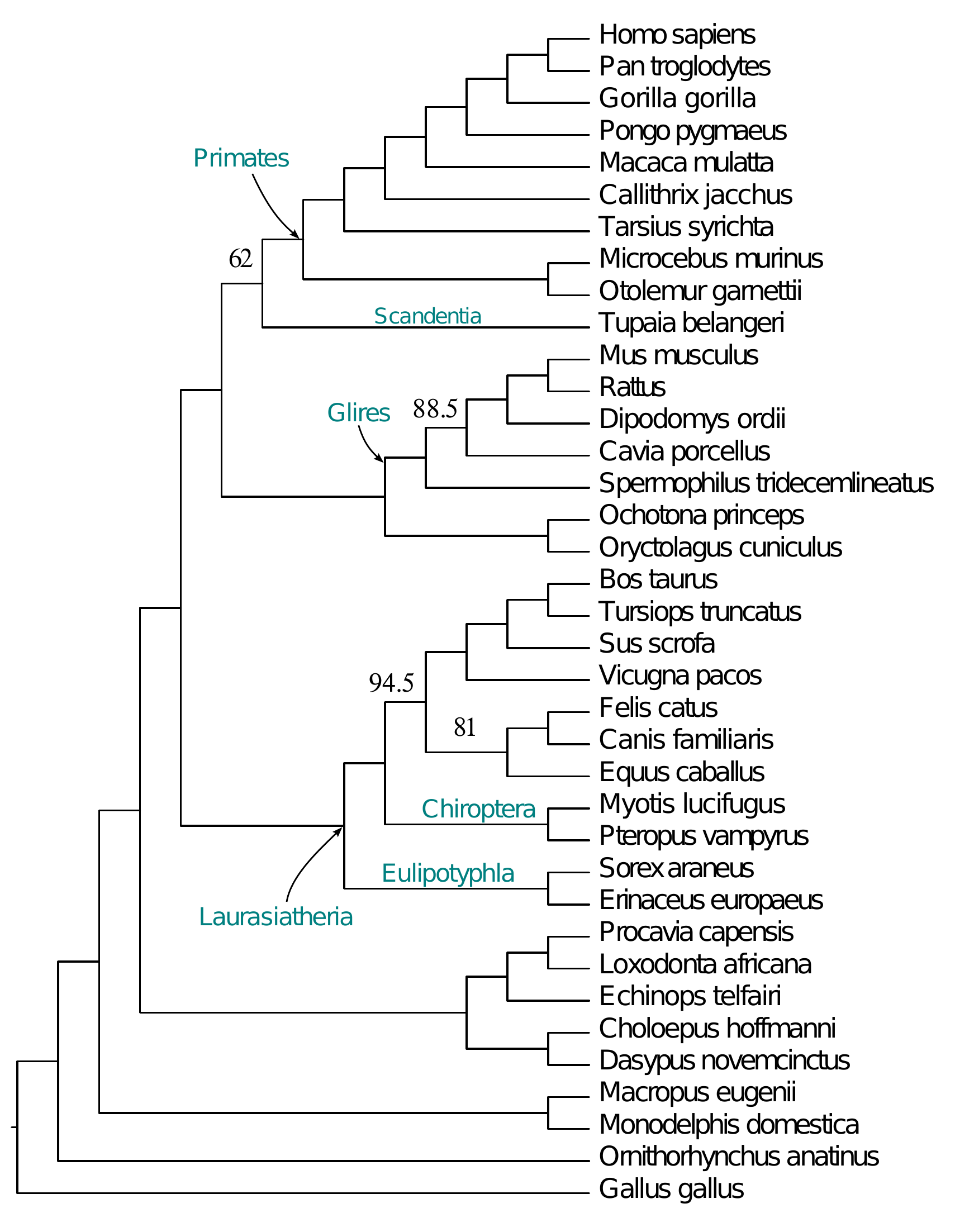}
\end{center}
\caption{Species trees estimated by unbinned MP-EST using MLBS for mammalian biological datasets. Branches without
designation have 100\% support. We used 75\% bootstrap support threshold for binning. We estimated the supergene trees using fully partitioned analyses.
} \label{fig:mammal-bio-astral}
\end{figure}


\begin{figure}[!ht]
\begin{center}
\includegraphics[width=0.80\textwidth]{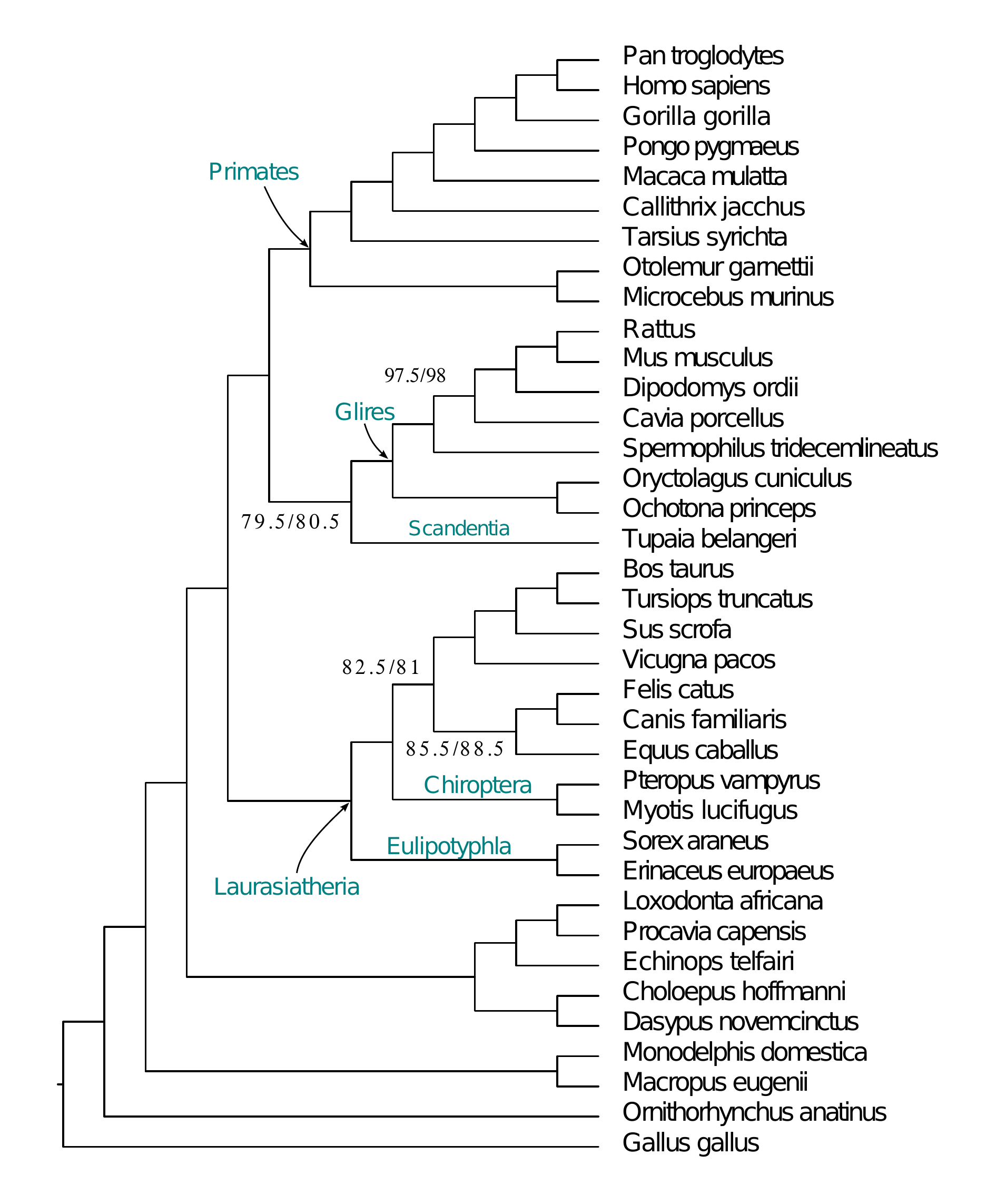}
\end{center}
\caption{Species trees estimated by binned (with and without weighting) MP-EST using MLBS for mammalian biological datasets. Binned and unbinned ASTRAL returned identical topology. The branches on this tree are labeled
with two support values side by side: the first one corresponds to unweighted binning and the next one
corresponds to weighted binning. Branches without
designation have 100\% support. We used 75\% bootstrap support threshold for binning. Supergene trees were estimated using fully partitioned analyses.
} \label{fig:mammal-bio-astral}
\end{figure}


\begin{figure}[!ht]
\begin{center}
\includegraphics[width=0.84\textwidth]{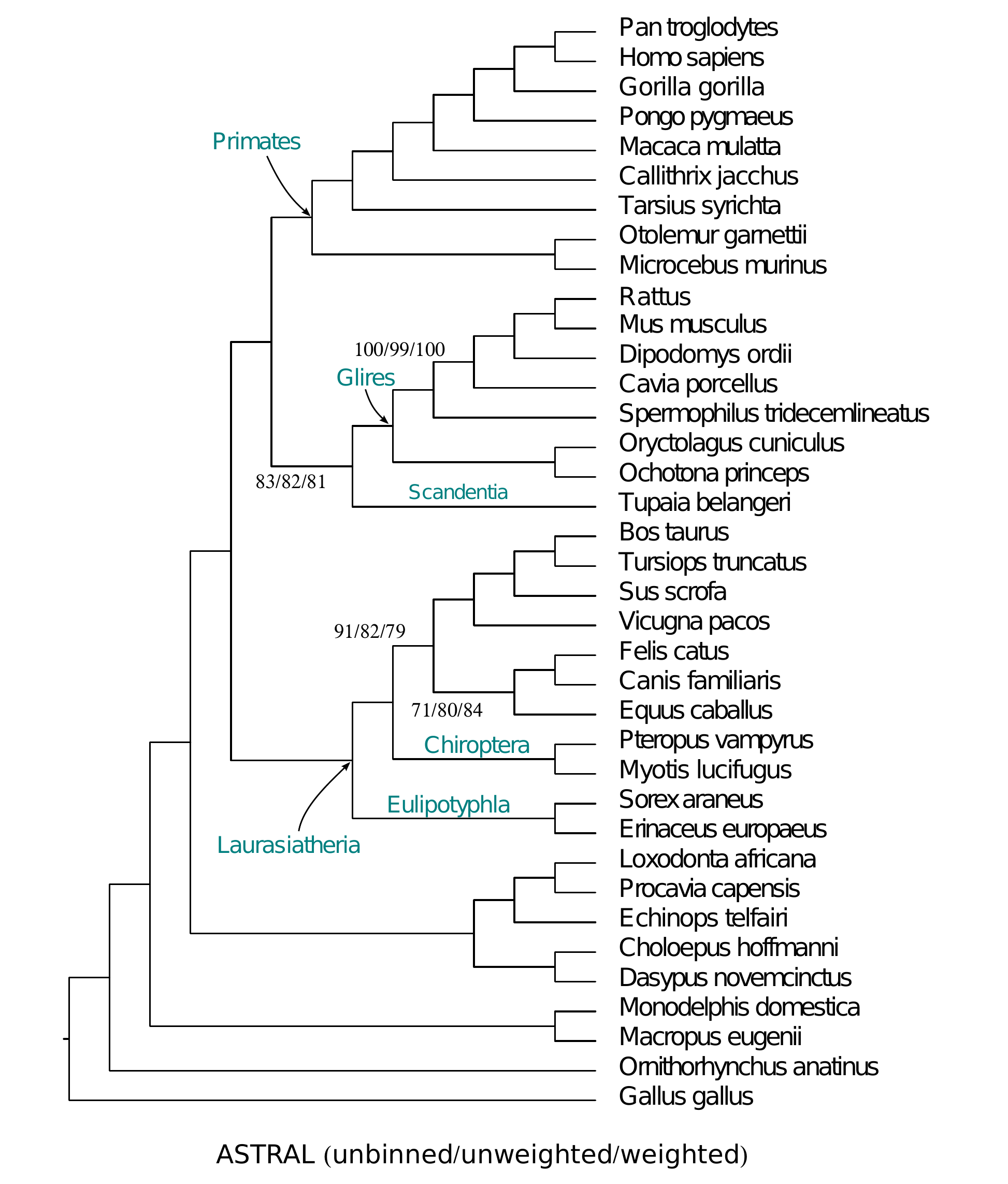}
\end{center}
\caption{Species trees estimated by unbinned and binned (with and without weighting) ASTRAL using MLBS for mammalian biological datasets. Binned and unbinned ASTRAL returned identical topology. The branches on this tree are labeled
with three support values side by side: the first one corresponds to unbinned ASTRAL, the next one
corresponds to unweighted binning, and the last one is for weighted binning. Branches without
designation have 100\% support. We used 75\% bootstrap support threshold for binning. Supergene trees were estimated using fully partitioned analyses.
} \label{fig:mammal-bio-astral}
\end{figure}


\begin{figure}[!ht]
\begin{center}
{\bf a) Control files used for MCcoal simulations (MCcoal.ctl): }\\\vspace{6pt}
\begin{verbatim}
SimulatedData.txt
9823126266
15 A B C D E F G H I J K L M N O
  1 1 1 1 1 1 1 1 1 1 1 1 1 1 1
((((((((((((((A #.05,B #.05):0.005 #.05,C #.05):0.01 #.05, D #.05):0.015 
#.05,E #.05):0.02 #.05,F #.05):0.025 #.05,G #.05):0.03 #.05,H #.05):0.035 
#.05,I #.05):0.04 #.05,J #.05):0.045 #.05,K #.05):0.05 #.05,L #.05):0.055 
#.05,M #.05):0.06 #.05,N #.05):0.065 #.05,O #.05):0.565 #.05;
\end{verbatim}

{\bf b) Command used to run MCcoal }\\\vspace{6pt}

\begin{verbatim}
   printf "10000 1000" PATH_TO_MCCOAL/MCcoal
\end{verbatim}

{\bf c) Commands to run bppseqgen }\\\vspace{6pt}

\begin{verbatim}
mkdir allTrees; 
split -a 4 -l 1 out.trees;
for i in x* ; do mv $i  allTrees/; done
for i in allTrees/x* ; do 
 bppseqgen number_of_sites=1000 input.tree.file=$i param=opts output.sequence.file=$i".fasta"
done
\end{verbatim}

{\bf d) GTR+$\Gamma$ model parameters }\\\vspace{6pt}

\begin{verbatim}
model = GTR(a=1.062409952497, b=0.133307705766, c=0.195517800882, 
d=0.223514845018, e=0.294405416545, 
theta=0.469075709819, theta1=0.558949940165, theta2=0.488093447144)
rate_distribution = Gamma(n=4, alpha=0.370209777709)
 \end{verbatim}
\end{center}
\caption{{\bf Simulation parameters and commands for 15-taxon dataset}.
Gene trees were simulated using MCcoal, with control files given here (a) and the command 
provided (b). The control files define the species tree, which is in the caterpillar form.
Running MCcoal simulated 10,000 gene trees, 
which we divided into 10 replicates of 1000 genes or 100 genes.
For each true gene tree, we then simulated alignments using
 bppseqgen \cite{Dutheil2008}, using the command given in (c).
 Here, the file ``opts'' is the same file we used in \cite{MirarabStatBinning} 
 and defines parameters given in (d).}
\end{figure}


\begin{figure}[!ht]
\begin{center}
{\bf a) Command used for SimPhy simulations: }\\\vspace{6pt}
\begin{verbatim}
for t in 1800000 200000; do
  b=0$(echo "scale=9; 1 / $t / 5"|bc -l}
  simphy -RS 20 -RL U:1000,1000 -RG 1 -ST U:$t,$t -SB U:$b,$b \\
              -SI U:1,1 -SL U:10,10 -CP U:400000,400000 -HS L:1.5,1 -HL L:1.2,1\\
              -HG L:1.4,1 -CU E:10000000 -SO U:1,1 -OD 1 -OR 0 -V 3  -CS 293745\\
              -O model.10.$t.$b |grep -E "[:-]"| tee log.10.$t.$b;
  for r in `ls -d model.$sp.$t.$b/*`; do
     sed -i -e "s/_0_0//g" $r/g_trees*.trees;
  done
done
\end{verbatim}

{\bf b) Parameter settings for SimPhy: }\\\vspace{6pt}
\begin{tabular}{llll}
\hline
Arg.&Description&Value&Notes\\
\hline
ST&maximum tree length& 200K or 1.8M& \\
SB& birth rates &0.000001 or 0.000000111&\\
SI& number of individuals per species&1&\\
SL& number of leaves& 10 &\\
P&global population sizes&400000 &\\
HS&Species-specific branch rate heterogeneity modifiers& Log normal (1.5,1)& \\
 HL&Locus-specific rate heterogeneity modifiers& Log normal (1.2,1) &\\
HG &Gene-tree-branch-specific rate heterogeneity modifiers& Log normal (1.4,1)&\\
 U & Global substitution rate& Exponential (10000000) &\\
 SO & Outgroup branch length relative to half the tree length& 1 &\\
\hline
\end{tabular}
\vspace{26pt}

{\bf c) Indelible GTR+$\Gamma$ model parameters }\\\vspace{6pt}
Base frequencies $\sim Dirichlet(36,26,28,32)$\\
GTR Matrix $\sim  Dirichlet(16,3,5,5,6,15)$\\
Rate parameter ($\alpha$) $\sim Exponential(1.2)$ trimmed at 0.1 from below
\end{center}

\caption{{\bf Simulation parameters and commands for 10-taxon dataset}.
Gene trees were simulated using SimPhy with commands given here (a)
which includes all the parameter settings (important parameters are listed in panel (b)). 
The maximum tree length parameter is set to either 
1.8M or 200K to produce two different model conditions. 
The speciation rate parameter is adjusted based on the maximum tree length
so that maximum rate multiplied by speciation rate is always 0.2
(thus, rate is 0.000000111 and 0.000001 for 1.8M and 200K respectively).
We used a perl script available at \url{http://www.cs.utexas.edu/~phylo/datasets/weighted-binning-datasets.html}
to draw parameters for the Indelible simulations.
Gene length is set to 1000 for all genes, 
but sequences are trimmed to their first 100bp in this study. 
For GTR+$\Gamma$ parameters, 
we use a set of hyper parameters (estimated from real datasets) to
drawn different parameter values for each gene in each replicate.
Hyper parameters for base frequencies, GTR matrix, and the rate 
parameter ($\alpha$) are shown in panel (c). 
These hyperparameters were calculated using maximum likelihood
estimation form a collection of three large scale multi-locus
datasets: 1KP dataset \cite{1kp}, Song et al Mammalian dataset \cite{Song2012},
and Avian phylogenomics dataset \cite{JarvisScience2014}. The base values
used for this maximum likelihood estimation and the
corresponding scripts are available at 
\url{http://www.cs.utexas.edu/~phylo/software/astral/} “(under the first bullet; i.e., estimate-parameters.r).
Note that for the shape frequency, $\alpha$, we use a heavy-tailed 
distribution, but to avoid unrealistic settings, 
values below 0.1 are discarded. 
}
\end{figure}

\end{document}